\documentclass[conference,10pt]{IEEEtran}
\pdfoutput=1
\usepackage[caption=false,font=footnotesize]{subfig}
\usepackage[utf8x]{inputenc}
\usepackage[T1]{fontenc}
\usepackage[english]{babel}
\usepackage[numbers,square,sort&compress]{natbib}
\usepackage{graphicx}
\usepackage{xspace}
\usepackage{verbatim}
\usepackage{url}
\usepackage[colorlinks=false]{hyperref} \usepackage{fancyvrb}
\usepackage{verbatim}
\usepackage{relsize}
\usepackage{listings}
\usepackage{booktabs} \usepackage{amssymb} \usepackage{pifont}
\usepackage{amsmath} 
\usepackage[usenames,dvipsnames,svgnames,table]{xcolor}
\usepackage{longtable}
\usepackage{caption}
\usepackage{multirow}
\usepackage{arydshln} 
\usepackage{flushend} 
\usepackage{tikz}
\newcommand*\circledb[1]{\tikz[baseline=(char.base)]{
    \node[shape=circle,draw,inner sep=1pt,fill=black] (char) {\textcolor{white}{\footnotesize #1}};}}

\newcommand*\circledw[1]{\tikz[baseline=(char.base)]{
    \node[shape=circle,draw,inner sep=1pt] (char) {\footnotesize #1};}}

\def\ifmonospace{\ifdim\fontdimen3\font=0pt }
\def\C++{\ifmonospace \C++ \else C\kern-.1167em\raise.25ex\hbox{\smaller{++}}\fi\xspace}

\def\lib\C++{\ifmonospace lib\C++ \else libc\kern-.1167em\raise.25ex\hbox{\smaller{++}} \fi\xspace}

\newcommand{\coolname}{CATT\xspace}
\newcommand{\coolnameBoot}{\mbox{B-CATT}\xspace}
\newcommand{\coolnameGeneric}{\mbox{G-CATT}\xspace}
\newcommand{\coolnameBoth}{\mbox{B/G-CATT}\xspace}
\newcommand{\domain}{security domain\xspace}
\newcommand{\domains}{security domains\xspace}

 \newcommand{\linuxkernelversion}{4.6\xspace}
\newcommand{\androidkernelversion}{4.4\xspace}

\newcommand{\cmark}{\ding{51}}\newcommand{\xmark}{\ding{55}}

\begin{document}
\title{\bf CAn't Touch This: Practical and Generic Software-only Defenses Against Rowhammer Attacks}

\author{\IEEEauthorblockN{
    Ferdinand Brasser,\IEEEauthorrefmark{1}
    Lucas Davi,\IEEEauthorrefmark{2}
    David Gens,\IEEEauthorrefmark{1}
    Christopher Liebchen,\IEEEauthorrefmark{1}
    Ahmad-Reza Sadeghi\IEEEauthorrefmark{1}
  }
  \IEEEauthorblockA{
    \IEEEauthorrefmark{1}CYSEC/Technische Universit\"at Darmstadt, Germany.\\\texttt{\{ferdinand.brasser,david.gens,christopher.liebchen,ahmad.sadeghi\}@trust.tu-darmstadt.de}\\
    \IEEEauthorrefmark{2}University of Duisburg-Essen, Germany. \texttt{lucas.davi@wiwinf.uni-due.de} \\
  }
}

\maketitle

\begin{abstract}
Rowhammer is a hardware bug that can be exploited to implement privilege escalation and remote code execution attacks. 
Previous proposals on rowhammer mitigation either require hardware changes or follow heuristic-based approaches (based on CPU performance counters).
To date, there exists no instant protection against rowhammer attacks on legacy systems.

\vspace{-5px}
In this paper, we present the design and implementation of two practical and efficient software-only defenses against rowhammer attacks.
Our defenses prevent the attacker from leveraging rowhammer to corrupt physically co-located data in memory that is owned by a different system entity.
Our first defense, \coolnameBoot, extends the system bootloader to disable vulnerable physical memory.
\coolnameBoot is highly practical, does not require changes to the operating system, and can be deployed on virtually all x86-based systems.
While \coolnameBoot is able to stop all known rowhammer attacks, it does not yet tackle the fundamental problem of missing memory isolation in physical memory.
To address this problem, we introduce our second defense \coolnameGeneric, a generic solution that extends the physical memory allocator of the OS to physically isolate the memory of different system entities (e.g., kernel and user space).

\vspace{-5px}
As proof of concept, we implemented \coolnameBoot on x86, and our generic defense, \coolnameGeneric, on x86 and ARM to mitigate rowhammer-based kernel exploits. Our extensive evaluation shows that both mitigation schemes
(i)~can stop available real-world rowhammer attacks,
(ii)~impose virtually no run-time overhead for common user and kernel benchmarks as well as commonly used applications, and
(iii)~do not affect the stability of the overall system.
\end{abstract}
\vspace{-7px}

\section{Introduction}
CPU-enforced memory protection is fundamental to modern computer security: for each memory access request, the CPU verifies whether this request meets the memory access policy. However, the infamous rowhammer attack~\cite{rowhammer-paper} undermines this access control model by exploiting a hardware fault (triggered through software) to flip targeted bits in memory. The cause for this hardware fault is due to the tremendous density increase of memory cells in modern DRAM chips, allowing electrical charge of one memory cell to affect that of an adjacent memory cell. 
\label{ref:refresh}
Unfortunately, increased refresh rates of DRAM modules -- as suggested by some hardware manufacturers -- cannot eliminate this effect~\cite{anvil}.
In fact, the fault appeared as a surprise to hardware manufacturers, simply because it does not appear during normal system operation. Rowhammer attacks repetitively read (\emph{hammer}) from the same physical memory address in very short time intervals which eventually leads to a bit flip in a physically co-located memory cell. 
\vspace{0.3cm}

\noindent \textbf{Rowhammer Attack Diversity.} Although it might seem that single bit flips are not per-se dangerous, recent attacks demonstrate that rowhammer can be used to undermine access control policies and manipulate data in various ways.
In particular, it allows for tampering with the isolation between user and kernel mode~\cite{rowhammer-exploit-google}. For this, a malicious user-mode application locates vulnerable memory cells and forces the operating system to fill the physical memory with page-table entries (PTEs), i.e., entries that define access policies to memory pages.
Manipulating one PTE by means of a bit flip allows the malicious application to alter memory access policies, building a custom page table hierarchy, and finally assigning kernel permissions to a user-mode memory page.
Rowhammer attacks have made use of specific CPU instructions to force DRAM access and avoid cache effects. However, prohibiting applications from executing these instructions, as suggested in~\cite{rowhammer-exploit-google},
is ineffective because recent rowhammer attacks do no longer depend on special instructions~\cite{anvil}.
As such, rowhammer has become a versatile attack technique allowing compromise of co-located virtual machines~\cite{rowhammer-vm1,rowhammer-vm2}, and enabling sophisticated control-flow hijacking attacks~\cite{stackdefiler, coop, jitrop} without requiring memory corruption bugs~\cite{rowhammer-exploit-google, rowhammer-browser, rowhammerjs}. Lastly, a recent attack, called Drammer~\cite{rowhammer-arm}, demonstrates that rowhammer is not limited to x86-based systems but also applies to mobile devices running ARM processors.

\noindent \textbf{Rowhammer Mitigation.}
The common belief is that the rowhammer fault cannot be fixed by means of any software update, but requires production and deployment of redesigned DRAM modules. Hence, existing legacy systems will remain vulnerable for many years, if not forever.
An initial defense approach performed through a BIOS update was unsuccessful as it only slightly increased the difficulty to conduct the attack~\cite{rowhammer-exploit-google}. The only other software-based mitigation of rowhammer, we are aware of, is a heuristic-based approach that relies on hardware performance counters~\cite{anvil}. However, it induces a worst-case overhead of 8\% and suffers from false positives which impedes its deployment in practice.

\noindent \textbf{Goals and Contributions.}
The goal of this paper is to develop the first practical and generic software-based defenses against rowhammer attacks that can instantly protect existing vulnerable legacy systems without suffering from any performance overhead and false positives. 
Intuitively, a defense against rowhammer needs to prevent the dangerous bit flips by disabling vulnerable memory pages. Kim et al.~\cite{rowhammer-paper} were first discussing this idea, however, they conclude that this mitigation will generate very high overhead and significantly reduce the amount of available memory. To validate this defense approach, we developed our first mitigation scheme, called \coolnameBoot,\footnote{The name \coolnameBoot is composed of two parts: B refers to our \textbf{B}ootloader based solution, CATT abbreviates \textbf{CA}n't \textbf{T}ouch \textbf{T}his.} that systematically scans the memory to identify and blacklist vulnerable memory pages. We implemented \coolnameBoot as a bootloader extension to support and protect a wide range of operating systems. In contrast, to the conclusion drawn in~\cite{rowhammer-paper}, we show that \coolnameBoot is highly efficient and prevents all existing rowhammer attacks. On the other hand, it is not yet known whether memory vulnerable to bit flips changes over time or under different experiment conditions (e.g., temperature, methodology to identify bit flips). This motivated us to develop a second and more generic mitigation scheme, called \coolnameGeneric,\footnote{G-CATT is short for \textbf{G}eneric CATT.} that does not aim to prevent bit flips but rather remove the dangerous effects (i.e., exploitation) of bit flips. This is achieved by limiting bit flips to memory pages that are already in the address space of the malicious application, i.e., memory pages that are per-se untrusted. For this, we extend the operating system kernel to enforce a strong physical isolation of different system entities, e.g., user and kernel space.

To summarize, our main contributions are:

\begin{itemize}
	\item We present practical software-based defenses against rowhammer. In contrast to existing solutions, our defenses require no hardware changes~\cite{rowhammer-paper}, do not deploy unreliable heuristics~\cite{anvil}, and still allow legacy applications to execute instructions that are believed to alleviate rowhammer attacks~\cite{rowhammer-exploit-google}.

\item We systematically revisit the practicality and effectiveness of an intuitive mitigation approach that proposes marking memory, vulnerable to rowhammer, unavailable to the system. We also developed a prototype, \coolnameBoot, which implements this approach for the first time.  

	\item We propose a generic enforcement mechanism for operating system kernels, \coolnameGeneric, to mitigate rowhammer attacks.
	Our design ensures that the attacker can only flip bits in memory that is already under her control.
	
	\item We present two prototype implementations, one for the Linux kernel version~\linuxkernelversion and one for the GRUB bootloader, and demonstrate their effectiveness in mitigating all previously presented rowhammer attacks~\cite{rowhammer-exploit-google, rowhammerjs}.
\label{ref:armintro}

	\item We successfully applied our Linux kernel patch for \coolnameGeneric to the Android version~\androidkernelversion for Google's Nexus devices. This allows us to also mitigate Drammer~\cite{rowhammer-arm}, a recent rowhammer-based privilege escalation exploit on ARM.
	
	\item We extensively evaluate the performance, robustness and security of our defense approaches against rowhammer attacks to demonstrate the effectiveness and high practicality of both \coolnameBoot~and \coolnameGeneric.
	In particular, our performance measurements indicate no computational overhead for common user and kernel benchmarks. 
\end{itemize}

\label{ref:focus-kexpl}
Note that recent research demonstrates a variety of exploitation strategies that rely on rowhammer~\cite{rowhammer-browser,rowhammer-vm1,rowhammer-vm2}.
While these attacks show the versatility of rowhammer, they are not always practical. For example, attacking the browser with rowhammer can take up to 45 minutes~\cite{rowhammer-browser}, and attacking a virtual machine may require a couple of days of preparation time~\cite{rowhammer-vm1}. Hence, our proof-of-concept implementation of \coolnameGeneric aims to mitigate  practical rowhammer attacks that escalate the privileges to the superuser root by manipulating the kernel page tables~\cite{rowhammer-exploit-google,rowhammer-arm}. We discuss further deployment cases in Section~\ref{sec:disc:gcatt}.
\begin{figure}[tp!]
  \centering
  \includegraphics[width=\linewidth]{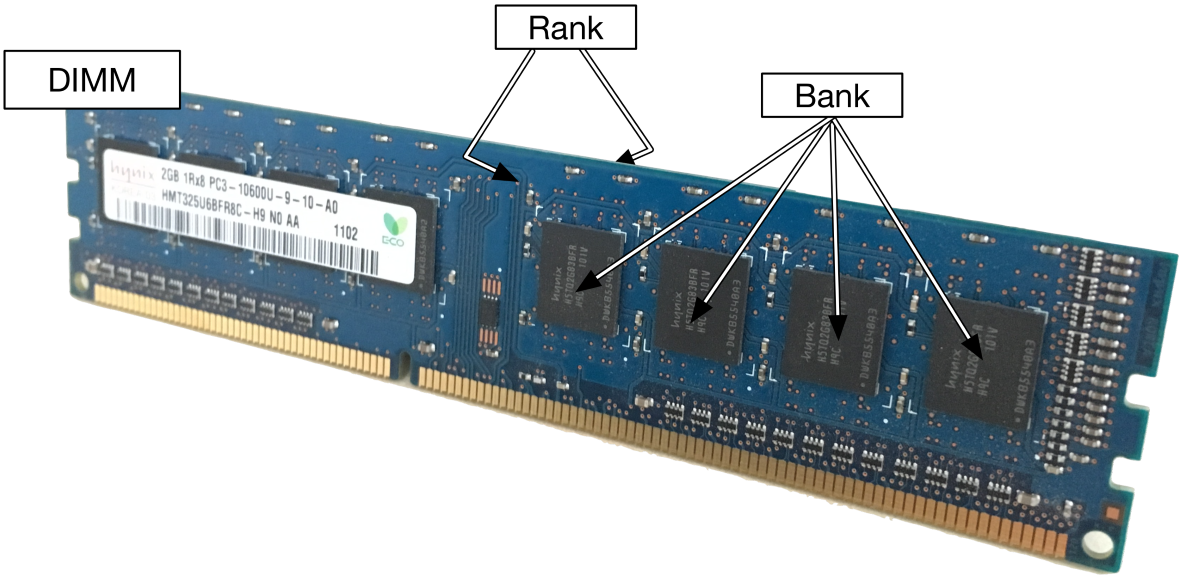}
    \caption{Organization of a DRAM module.}
  \label{fig:dram}
\end{figure}

\section{Background}
In this section we provide the basic background knowledge necessary for understanding the remainder of this paper.

\subsection{Dynamic Random Access Memory (DRAM)}
\label{sec:background:dram-mapping}
\begin{figure}[tp!]
  \centering
  \includegraphics[width=.8\linewidth]{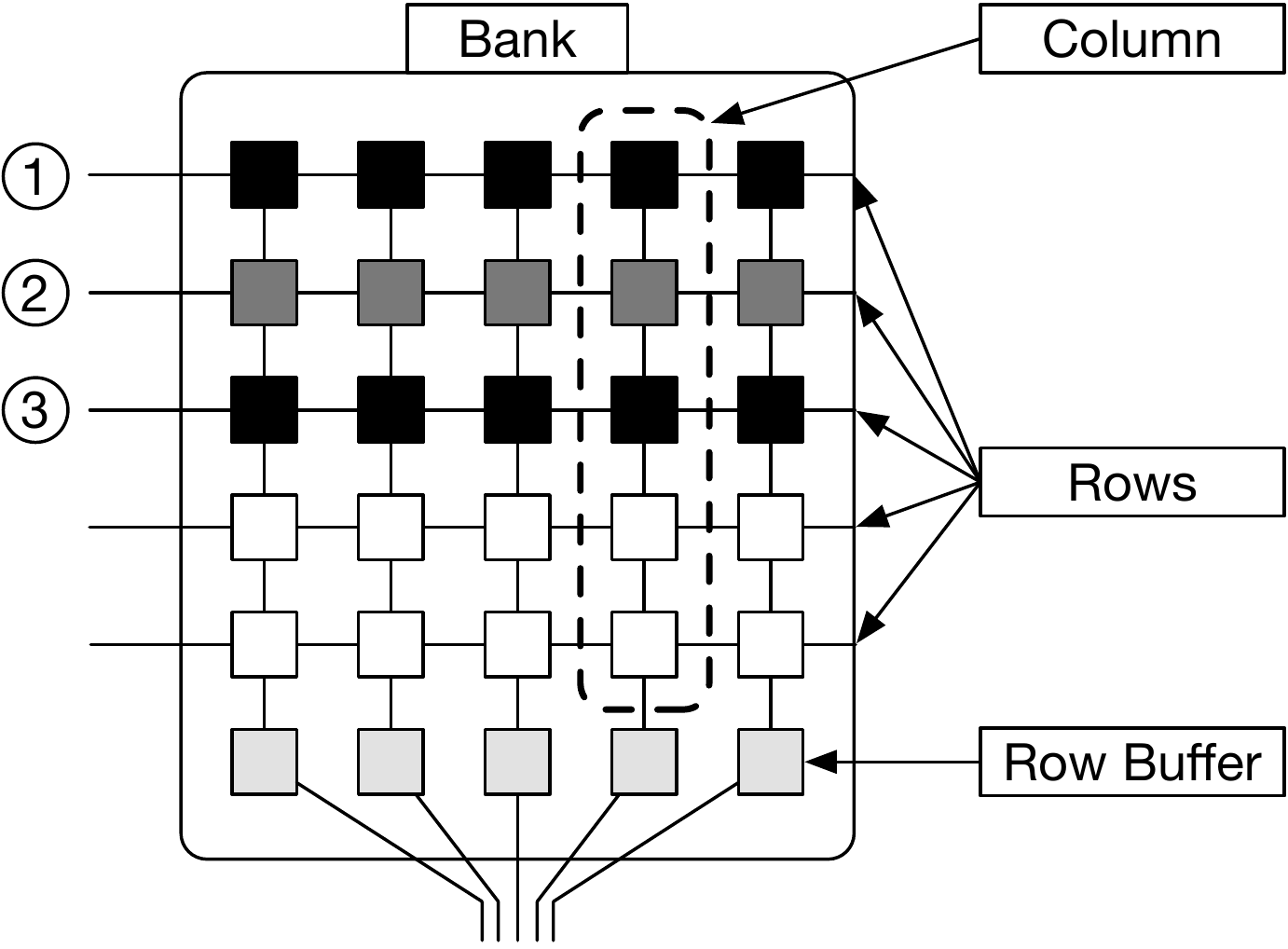}
    \caption{Organization of a Bank.}
  \label{fig:dram2}
\end{figure}

A DRAM module, as shown in Figure~\ref{fig:dram}, is structured hierarchically.
The hardware module is called Dual Inline Memory Module (DIMM), which is physically connected through a channel to the memory controller.
Modern desktop systems usually feature two channels facilitating parallel accesses to memory. The DIMM can be divided into one or two ranks corresponding to its front- and backside.
Each rank contains multiple banks which are the chips that contain the memory cells.
Each bank is organized in columns and rows, as shown in Figure~\ref{fig:dram2}.

An individual memory cell consists of a capacitor and a transistor.
To store a bit in a memory cell, the capacitor is electrically charged.
By reading a bit from a memory cell, the cell is discharged, i.e., read operations are destructive. 
To prevent information loss, read operations also trigger a process that writes the bit back to the cell.
A read operation always reads out the bits from a whole row, and the result is first saved in the row buffer before it is then transferred to the memory controller.
The row buffer is also used to write back the content into the row of memory cells to restore their content.

It is noteworthy to mention that there exists a non-linear mapping between physical memory address and the rank-bank-row on the hardware module.
Consequently, two consecutive physical memory addresses can be mapped to memory cells that are located on different ranks, banks, or rows.
For example, on Intel Ivy Bridge CPUs the 20th bit of the physical address determines the rank.
As such, the consecutive physical addresses 0x2FFFFF and 0x300000 can be located on front and back side of the DIMM for this architecture.
The knowledge of the physical memory location  on the DIMM is important for both rowhammer attacks and defenses, since bit flips can only occur on the same bank.
For Intel processors, the exact mapping is not officially documented, but has been reverse engineered~\cite{drama,rowhammer-vm2}.

\subsection{Rowhammer Overview and Challenges}
\label{sec:background:rowhammer}

As mentioned before, memory access control is an essential building block of modern computer security, e.g., to achieve process isolation, isolation of kernel code, and manage read-write-execute permission on memory pages. Modern systems feature a variety of mechanisms to isolate memory, e.g., paging~\cite{intel-manual}, virtualization~\cite{amd-virt, intel-exe-only-support}, IOMMU~\cite{iommu}, and special execution modes like SGX~\cite{intel-manual} and SMM~\cite{intel-manual}.
 However, these mechanisms enforce their isolation through hardware that mediates the physical memory accesses (in most cases the CPU).
Hence, memory assigned to isolated entities can potentially be co-located in physical memory on the same bank.
Since a rowhammer attack induces bit flips in co-located memory cells, it provides a subtle way to launch a remote attack to undermine memory isolation.

Recently, various rowhammer-based attacks have been presented.
Specifically, rowhammer was utilized to undermine isolation of operating system and hypervisor code, and escape from application sandboxes leveraged in web browsers.
In the following, we describe the challenges and workflow of rowhammer attacks. A more elaborated discussion on real-world, rowhammer-based exploits will be provided in Section~\ref{sec:relatedwork}.

The rowhammer fault allows an attacker to influence the electrical charge of individual memory cells by activating neighboring memory cells.
In particular, Kim et al.~\cite{rowhammer-paper} demonstrate that repeatedly activating two rows separated by only one row, called \emph{aggressor rows} (\circledw{1} and~\circledw{3} in Figure~\ref{fig:dram2}), lead to a bit flip in the enclosed row~\circledw{2}, called \emph{victim row}. 
To do so, the attacker has to overcome the following challenges: (i)~undermine memory caches to directly perform repetitive reads on physical DRAM memory, and (ii)~gain access to memory co-located to data critical to memory isolation.

Overcoming challenge (i) is complicated because modern CPUs feature different levels of memory caches which mediate read and write access to physical memory.
Caches are important as processors are orders of magnitude faster than current DRAM hardware, turning memory accesses
into a bottleneck for applications~\cite{wulf1995hitting}.
Usually, caches are transparent to software, but many systems feature special instructions, e.g., \texttt{clflush} or \texttt{movnti} for x86~\cite{rowhammer-exploit-google,rowhammer-new-attack}, to undermine the cache. Further, caches can be undermined by using certain read-access patterns that force the cache to reload data from physical memory.
Such patterns exist, because CPU caches are much smaller than physical memory, and system engineers have to adopt an eviction strategy to effectively utilize caches.
Through alternating accesses to addresses which reside in the same cache line, the attacker can force the memory contents to be fetched from physical memory.

The attacker's second challenge (ii) is to achieve the physical memory constellation shown in Figure~\ref{fig:dram2}.
The attacker needs access to the aggressor rows in order to be able to activate (\emph{hammer}) them (rows \circledw{1} and~\circledw{3} in Figure~\ref{fig:dram2}).
Additionally, the victim row must contain data which is attackable by a bit flip (\circledw{2} in Figure~\ref{fig:dram2}).
Both conditions cannot be enforced by the attacker, however, he can achieve them with high probability through the following approaches. 
First, the attacker allocates memory hoping that the aggressor rows are contained in the allocated memory.
If the operating system maps the attacker's allocated memory to the physical memory containing the aggressor rows the attacker has achieved the first condition.
Since the attacker has no influence on the mapping between virtual memory and physical memory he cannot directly influence this step, but he can increase the probability by allocating large amounts of memory; also, the attacker can repeat this step until he succeeds.
Once the attacker has the aggressor rows under his control he releases all allocated memory except the parts containing the aggressor rows.
Next, the attacker needs the data he wants to manipulate to be stored in the victim row. 
Again, he cannot directly influence which data is stored in the physical memory and needs to resort to a probabilistic approach.
The attacker induces the creation of many copies of the victim data with the goal that one copy of the victim data will be placed in the victim row.
The attacker cannot directly verify whether the second step was successful, he has to execute the rowhammer attack and check if the attack is successful, otherwise he has to repeat the second step.

In~\cite{rowhammer-exploit-google}, the authors successfully implemented this approach to compromise the kernel from an unprivileged user process.
They gain control over the aggressor rows and then make the OS create huge amounts of page table entries with the goal of placing one page table entry in the victim row.
By flipping a bit in a page table entry the authors gain control over a subtree of the page tables allowing them to manipulate memory access control policies.

\section{Threat Model and Assumptions} \label{sec:threat-model}
Our threat model is in line with related work~\cite{rowhammer-exploit-google,rowhammer-new-attack, rowhammer-browser, rowhammerjs, rowhammer-vm1,rowhammer-vm2}: 
rowhammer-based attacks assume that the adversary is capable of executing malicious code in an isolated environment. Depending on the attack scenario, the isolated environment can be a JavaScript to launch browser-based attacks, a malicious user-mode application to construct kernel exploits, or an entire operating system for running attacks against co-located virtual machines in a cloud setting.
The goal of an adversary is to undermine the isolation. In other words, the attacker aims to break out of the established trust boundary by inducing bit flips in memory that does not belong to the isolated environment. Ultimately, an adversary gains a higher privilege level than originally assigned to the isolated environment.

Our threat model differentiates between \emph{reliable} and \emph{unreliable} bit flips. The former can be reproduced, e.g., across multiple reboots, the latter cannot be produced on every attempt to flip a bit through rowhammer.
Note that \emph{all} available rowhammer-based attacks depend on reliable bit flips. However, in this paper, we consider software-defenses against both types of bit flips.

\section{Overview and Design}
\label{sec:highlvl:gcatt}
          
\begin{figure*}[!t]
\centering
\subfloat[\coolnameBoot disables vulnerable memory pages]{\includegraphics[height=0.25\linewidth]{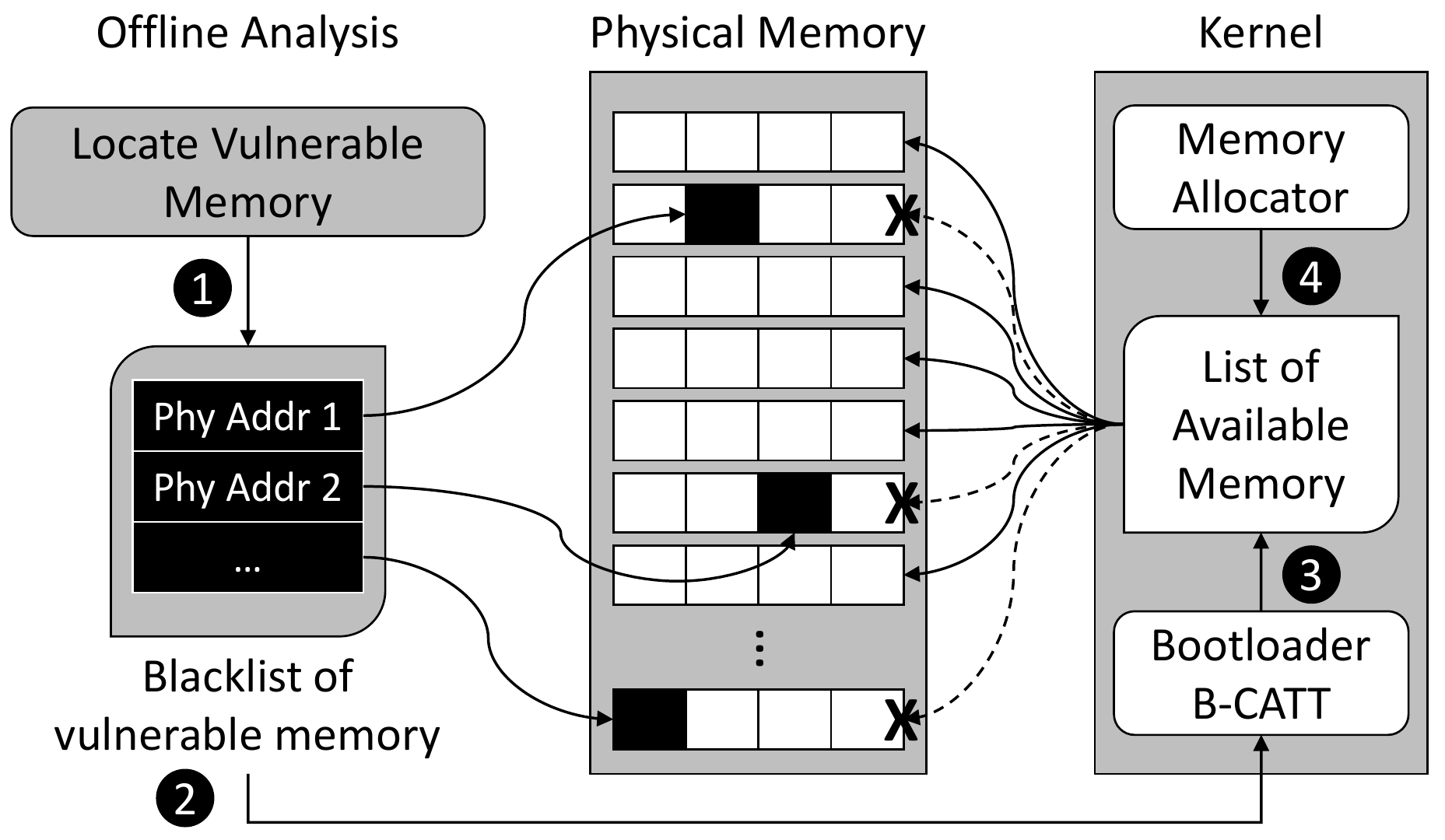}\label{fig_b-catt_design}}
\hfil
\subfloat[\coolnameGeneric constrains bit flips to the process' security domain]{\includegraphics[height=0.25\linewidth]{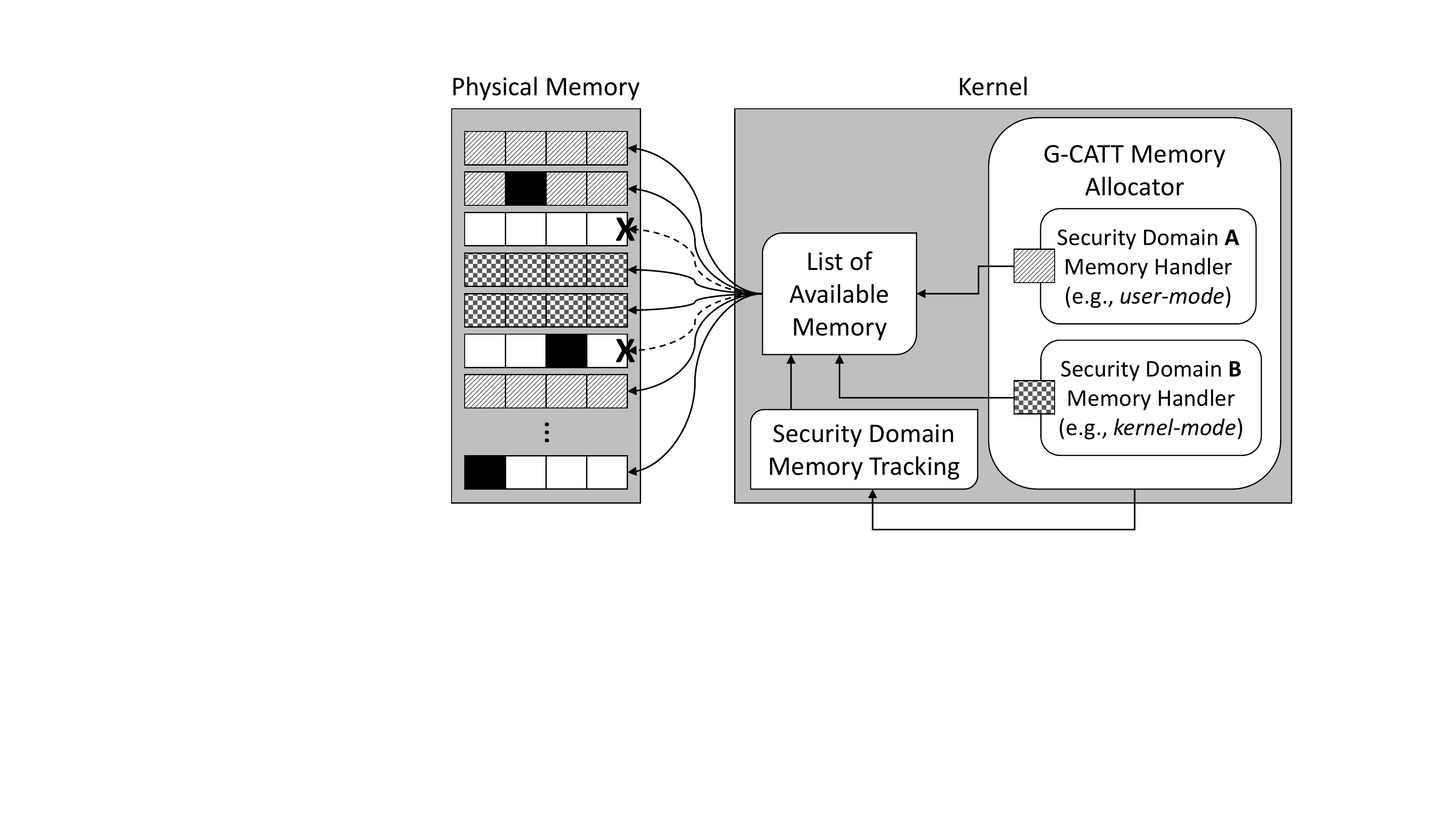}\label{fig_g-catt_design}}
\caption{Overview of our software-based defenses against rowhammer attacks.}
\label{fig_design}
\vspace*{-0.2cm}
\end{figure*}

In this section, we present the high-level idea and design of our practical software-based defenses against rowhammer attacks. Our first defense, dubbed \coolnameBoot, mitigates malicious \emph{bit flips} by marking memory vulnerable to rowhammer as unavailable at boot-time. In contrast, our second defense, called \coolnameGeneric, tackles the malicious \emph{effect} of rowhammer-induced bit flips by instrumenting the operating system's memory allocator to constrain bit flips to the boundary where the attacker's malicious code executes (cf. Section~\ref{sec:threat-model}). Both solutions \coolnameBoot and \coolnameGeneric are completely transparent to applications, and do not require any hardware changes. 
We provide a detailed description of our implementation of both solutions in Section~\ref{ref:limite820}.

\subsection{Boot-\coolname}
\label{sec:highlvl:bcatt}

Prior to launching a rowhammer attack, the attacker needs to probe the memory to locate vulnerable memory. 
Hence, a software-only defense needs to identify the vulnerable memory parts and mark them as unavailable. 
While simple in theory, it was commonly believed that this defense strategy is inefficient and significantly reduces the amount of available memory~\cite{rowhammer-paper}. 
To validate this, we developed \coolnameBoot, a new bootloader extension that locates and disables vulnerable memory parts (see Figure~\ref{fig_b-catt_design}). Implementing this defense strategy is challenging without inducing performance penalties. While we investigated different implementation strategies, we opted for a bootloader extension in order to protect a wide range of operating systems. That is, our bootloader-based approach does not require any changes to the operating system and is fully compatible with legacy systems.

In general, all bootloaders commit a list of available memory to the operating system~(\circledb{3}). This list is dynamically obtained during boot time using the system's firmware (e.g., BIOS/UEFI).
It is common that certain physical memory addresses cannot be used by the operating system, e.g., the memory reserved for hardware devices such as graphics or network cards.
\coolnameBoot extends the bootloader and hooks into this process. Specifically, it adds the physical addresses of vulnerable memory to the list of unavailable memory~(\circledb{2}).
To locate vulnerable addresses, we re-use techniques from existing rowhammer exploitation tools~\cite{rowhammerjs,rowhammer-memtest} and slightly adjust them for the purpose of our approach. Note that this analysis is performed offline prior the system is protected with \coolnameBoot~(\circledb{1}).
From the operating system's perspective, the vulnerable physical memory simply becomes unavailable~(\circledb{4}). Since handling unavailable memory regions is a core functionality of operating systems, \coolnameBoot adheres to common system functionality.

\subsection{Generic-\coolname}
Note that \coolnameBoot takes a snapshot of vulnerable memory addresses. Since there is no security guarantee 
that vulnerable memory changes over time or other memory addresses get vulnerable under different test conditions, 
we developed a second prototype that provides long-term protection against rowhammer attacks.
Our second defense, called \coolnameGeneric, follows a different and more generic defense strategy: it tolerates rowhammer-induced bit flips, but prevents bit flips from affecting memory belonging to higher-privileged \emph{\domains}, e.g., the operating system kernel or co-located virtual machines.
As discussed in Section~\ref{sec:background:rowhammer}, a rowhammer attack requires the adversary to bypass the CPU cache.
Further, the attacker must arrange the physical memory layout such that the targeted data is stored in a row that is physically adjacent to rows that are under the control of the attacker.
Hence, \coolnameGeneric ensures that memory between these two entities is physically separated by at least one row.\footnote{Kim et al.~\cite{rowhammer-paper} mention that the rowhammer fault does not only affect memory cells of directly adjacent rows, but also memory cells of rows that are next to the adjacent row. Although we did not encounter such cases in our experiments, \coolnameGeneric supports multiple row separation between adversary and victim data memory.}

To do so, \coolnameGeneric extends the physical memory allocator to partition the physical memory into \domains.
\coolnameGeneric supports a wide range of memory partitioning levels, i.e., the granularity of \domains can be tuned from coarse-grained levels such as the separation between user and kernel memory to fine-grained memory partition within a single process. 

Figure~\ref{fig_g-catt_design} illustrates the concept. Without \coolnameGeneric the attacker is able to craft a memory layout, where two aggressor rows enclose a victim row of a higher-privileged domain such as row \circledw{2} in Figure~\ref{fig:dram2}.
With \coolnameGeneric in place, the rows which are controlled by the attacker are grouped into the \domain~A, whereas memory belonging to higher-privileged entities resides with their own \domain (e.g., the \domain~B).
Both domains are physically separated by at least one row which will not be assigned to any \domain.

\noindent \textbf{Security Domains.}
The definition of \domains is critical to ensure the effectiveness of \coolnameGeneric. Judging from previous privilege escalation attacks, good candidates for such domains are kernel and user-mode, virtual machines, as well as plugins and scripts in web browsers. Privilege escalation attacks are popular and pose a severe threat to modern systems. In particular, the isolation of kernel and user-mode is critical and the most appealing attack target. If a user-space application gains kernel privileges, the adversary can typically compromise the entire system. 
Due to the importance of these attacks and the fact that rowhammer-based exploits on kernels are highly practical, we focus our proof-of-concept implementation on kernel and user-space isolation. That is, we define and maintain two security domains: a \domain for kernel memory allocations, and one \domain for user-mode memory allocations (see also Figure~\ref{fig_g-catt_design}). Note that \coolnameGeneric can be extended to support further \domains. For instance, in virtualized environments, virtual machines as well as the hypervisor can be treated as individual \domains. To prevent rowhammer attacks crossing the isolation boundaries of processes in user-mode, \coolnameGeneric would need to assign \domains at the granularity of processes. However, many \domains associated with many memory allocations and deallocations will eventually lead to inefficient usage of memory, i.e., memory fragmentation. If memory fragmentation is an issue, \coolnameGeneric could resort in only assigning a dedicated security domain for untrusted third-party processes. For in-process isolation, e.g., to prevent rowhammer-based sandboxing escapes in browsers~\cite{rowhammer-browser}, \coolnameGeneric needs to distinguish the allocations of trusted browser code and untrusted sandboxed code. Although we have not yet implemented \domains beyond kernel and user-mode, we believe that there no major conceptual obstacles in supporting \domains for each of the described scenarios.

\noindent \textbf{Challenges.}
The physical isolation of data raises the challenge of how to effectively isolate the memory of different system entities. To tackle this challenge, we first require knowledge of the mapping between physical addresses and memory banks.
Since an attacker can only corrupt data within one bank, but not across banks, \coolnameGeneric only has to ensure that \domains of different system entities are isolated within each bank.
However, as mentioned in Section~\ref{sec:background:dram-mapping}, hardware vendors do not specify the exact mapping between physical address and banks.
Fortunately, Pessl et al.~\cite{drama} and Xiao et al.~\cite{rowhammer-vm2} provide a methodology to reverse engineer the mapping.
For \coolnameGeneric, we use this methodology to discover the physical addresses of rows.

Lastly, we need to ensure that the physical memory management component is aware of the isolation policy. 
This is vital as the memory management components have to ensure that newly allocated memory is adjacent only to memory belonging to the same security domain. 
To tackle this challenge, we instrumented the memory allocator to keep track of the domain association of physical memory and serve memory requests by selecting free memory from different pools depending on the security domain of the requested memory.

\section{Implementation}
\label{sec:impl}
Both of our software-based defenses are based on modifications to low-level system software components, i.e., the bootloader (\coolnameBoot) and the operating system kernel (\coolnameGeneric).
We chose GRUB2 and Linux 4.6 as targets for our proof-of-concept implementations for two reasons:
(1) their source code is freely available, and
(2) both projects are widely used.
First, we briefly describe how \coolnameBoot blacklists rowhammer-affected pages.
Thereafter, we explain the implementation of \coolnameGeneric's policy enforcement mechanism in the Linux kernel which allows for the partitioning of physical memory into isolated \domains. Note that our prototype implementation of \coolnameBoot targets x86-based systems, whereas \coolnameGeneric targets both x86 and ARM-based systems. Until today, rowhammer attacks have been only demonstrated for these two prominent architectures. Furthermore, as mentioned before, our prototype implementations target rowhammer exploits that aim at violating the isolation between kernel and user-mode since these are the most prominent and practical types of rowhammer exploits.

\subsection{B-CATT}
\label{sec:impl:bcatt}
To mitigate the exploitation of the rowhammer hardware bug in legacy systems, \coolnameBoot features a blacklisting mechanism to prevent the operating system from using vulnerable pages.
Recall that Kim et al.~\cite{rowhammer-paper} concluded that blacklisting is not practical if the majority of rows are susceptible to rowhammer.
However, our evaluation (Section~\ref{sec:eval:rowhammertest:bcatt}) suggests that only a fraction of rows are vulnerable in practice. This is also in line with previous work~\cite{rowhammer-exploit-google}.
In fact, we noticed that blacklisting vulnerable memory is highly practical as it only requires a small extension to the bootloader (64 lines of code).
We exploit the fact that operating systems already feature a function that allows blacklisting certain physical memory addresses.
Specifically, x86 compatible operating systems detect usable memory through the so-called \texttt{e820 map}~\cite{e820}.
This map contains a list of usable physical memory. It can be obtained from the BIOS through the interrupt number 0xe820.
On modern systems, this map is not directly requested by the operating system. In contrast, the bootloader forwards the map to the operating system.
\label{ref:limite820}
To prevent rowhammer attacks, \coolnameBoot instruments the GRUB2 bootloader to add physical addresses of vulnerable memory to the \texttt{e820 map}.
Note that the original e820 specification only supports 128 entries. However, the EFI Platform Initialization specification -- which is also supported by GRUB2 -- does not limit the number of entries~\cite{intelEFIe820}.
In fact, GRUB2 first allocates a buffer to store both the original e820 map and additional entries.
Once it has stored the original e820 entries into this buffer, it passes a pointer to this buffer to the operating system.
Our \coolnameBoot patch increases the size of the allocated buffer to additionally fit the entries for the blacklisted pages.

\label{sec:highlvl:gcatt}
\begin{figure}[tp]
  \centering
  \includegraphics[width=\linewidth]{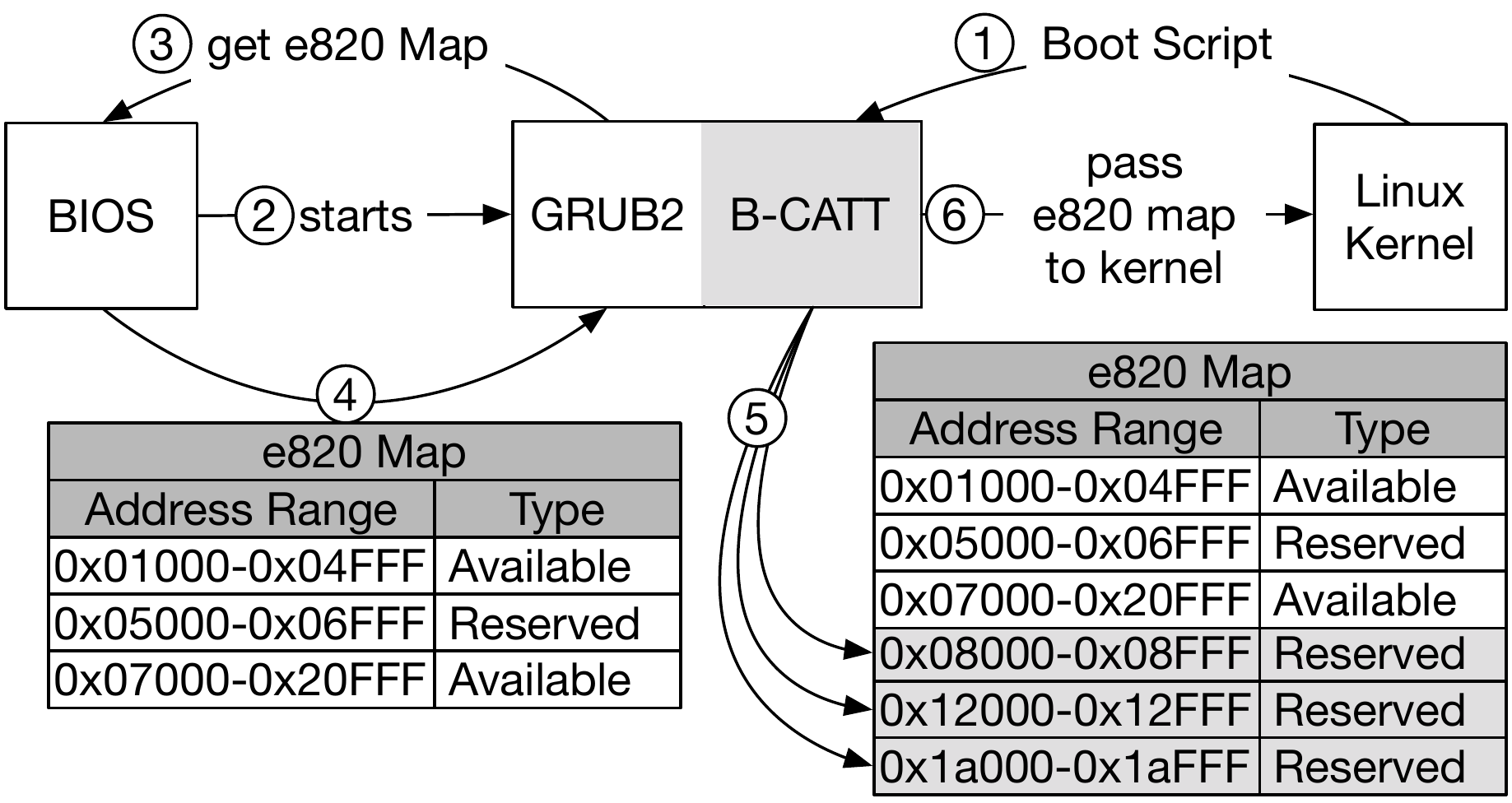}
    \caption{Workflow of \coolnameBoot}
  \label{fig:bcatt-detail}
\end{figure}

Figure~\ref{fig:bcatt-detail} depicts the detailed workflow of \coolnameBoot.
After upgrading the existing GRUB2 installation, the vulnerable memory rows are identified with the help of publicly available tools during an offline analysis~\cite{rowhammer-exploit-google, rowhammerjs, rowhammer-memtest}.
These programs scan the memory of the machine for vulnerable pages by allocating large amounts of memory and trying to produce bit flips.
We made some modifications, which we describe in detail in Section \ref{sec:rowhammer-test}.

Further, these rows are included as an environment variable in the boot configuration file of GRUB2 (Step~\circledw{1}).
Next, the BIOS starts GRUB2 which then requests the \texttt{e820 map} (Step~\circledw{2}--\circledw{4}).
Before GRUB2 starts the operating system, it reads out the environment variable and extends the \texttt{e820 map} (Step~\circledw{5}--\circledw{6}). 
Henceforth, the Linux kernel will not use the vulnerable memory thereby preventing the attacker from leveraging rowhammer-based attacks.

\subsection{G-CATT}
In our proof-of-concept implementation of \coolnameGeneric, we focus on hardening Linux against rowhammer-based attacks since its memory allocator is open-source.
We successfully applied the described changes to the x86-kernel version 4.6 and the Android kernel for Nexus devices in version 4.4.\label{ref:armimpl}

The basic idea underlying our generic software-based rowhammer defense is to physically separate rows 
that belong to different security domains. Operating systems are not per-se aware of the 
notions of cells and rows, but rather build memory management based on paging.
Commodity operating systems use paging to map virtual addresses to physical addresses.
The size of a page varies among architectures. On x86 and ARM, the page size is typically 4096 bytes (4K).
As we described in Section~\ref{sec:background:dram-mapping}, DRAM hardware consists of much smaller
units of memory, i.e., individual memory cells storing single bits.
Eight consecutive memory cells represent a byte, 4096 consecutive bytes a page frame, two to four page frames a row.
Hence, our implementation of \coolnameGeneric changes low-level components of the kernel to make
the operating system aware of the concept of memory rows. 

In the following, we describe how we map individual memory pages to domains, keep track of different domains, modify the physical memory allocator, and define partitioning policies for the system's DRAM hardware.

\subsubsection{Mapping Page Frames to Domains}
To be able to decide whether two pages belong to the same security domain we need to keep track of the security domain for each page frame.
Fortunately, the kernel already maintains meta data about each individual page frame.
More specifically, each individual page frame is associated with exactly one meta data object (\texttt{struct page}).
The kernel keeps a large array of these objects in memory.
Although these objects describe physical pages, this array is referred to as
\emph{virtual memory map}, or \texttt{vmemmap}.
The Page Frame Number (PFN) of a physical page is used as an offset into this array to determine the corresponding \texttt{struct page} object.
To be able to associate a page frame with a security domain, we extend the definition of \texttt{struct page} to include a field that encodes the security domain.
Since our prototype implementation targets rowhammer attacks that aim at violating the separation of kernel and user-space, we encode 
security domain 0 for kernel-space, and 1 for user-space.
For preventing rowhammer attacks crossing the process boundary, \coolnameGeneric could exploit the process identification number
(PID) to distinguish security domains. As such, each process will be treated as a separate security domain, and the kernel will be associated with its own security domain.

\subsubsection{Tracking Security Domains}
\label{sec:impl:processisolation}
The extension of the page frame meta data objects enables us to assign pages to security domains.
However, this assignment is dynamic and changes over time.
In particular, a page frame may be requested, allocated, and used by one domain, after it has been freed by another domain.
Note that this does not violate our security guarantees, but is necessary for the system to manage physical memory dynamically.
Yet, we need to ensure that page frames being reallocated continue to obey our security policy.
Therefore, we reset the security domain upon freeing a page. 

Upon memory allocation, \coolnameGeneric needs to correctly set the security domain of the new page. 
To do so, we require information about the requesting domain. 
For our case, where we aim at separating kernel and user-space domains, \coolnameGeneric utilizes 
the call site information, which is propagated to the memory allocator by default.
Specifically, each allocation request passes a range of flags to the page allocator.
These flags encode whether an allocation is intended for the kernel or the user-space.
We leverage this information and separate the two domains by setting the
domain field of the respective page frame. 

When processes request memory, the kernel initially only creates a virtual mapping without providing actual physical page frames for the process.
Instead, it only assigns physical memory on demand, i.e., when the requesting process accesses the virtual mapping a page fault is triggered.
Thereafter, the kernel invokes the physical page allocator to search for usable pages and installs them under the virtual address the process attempted to access.
We modified the page fault handler, which initiates the allocation of a new page, to pass information about the security domain to the page allocator.
Next, the page is allocated according to our policy and sets the domain field of the page frame's meta data object to the security domain of the interrupted process.

\subsubsection{Modifying the Physical Page Allocator}
The Linux kernel uses different levels of abstraction for different memory allocation tasks.
The physical page allocator, or \emph{zoned buddy allocator}
is the main low-level facility handling physical page allocations.
It exports its interfaces through functions such as \texttt{alloc\_pages}, which can be used by other kernel components to request physical pages.
In contrast to higher-level allocators, the buddy allocator only allows for allocating sets of memory pages with a cardinality which can be expressed as a power of two (this is referred to as the \emph{order} of the allocation).
Hence, the buddy allocator's smallest level of granularity is a single memory page.
We modify the implementation of the physical page allocator in the kernel to include a mechanism for separating and isolating allocated pages according to the security domain of the origin of the allocation request.
In particular, the page allocator already performs maintenance checks on free pages.
We extend these maintenance checks to add our partitioning policy before the allocator returns a physical page.
If this check fails, the page allocator is not allowed to return the page in question, but has to continue its search for another free page.

\subsubsection{Defining DRAM Partitioning Policies}
Separating and isolating different security domains is essential to our proposed mitigation.
For this reason, we incorporate detailed knowledge about the platform and its DRAM hardware configuration into our policy implementation.
While our policy implementation for a target system largely depends on its architecture and memory configuration, this does not represent a fundamental limitation.
Indeed, independent research~\cite{drama,rowhammer-vm2} has provided the architectural details for the most prevalent architectures, i.e.,
it shows that the physical address to DRAM mapping can be reverse engineered automatically for undocumented architectures.
Hence, it is possible to develop similar policy implementations for architectures and memory configurations beyond x86 and ARM.
We build on this prior research and leverage the physical address to DRAM mapping information to enforce strict physical isolation.
In the following, we describe our implementation of the partitioning strategy for isolating kernel and user-space, as well as providing an implementation strategy for isolating individual user processes.

\noindent \textbf{Kernel-User Isolation.} To achieve physical separation of user and kernel space we adopt the following strategy: we divide each bank into a top and a bottom part, with a separating row in-between.
Page frames for one domain are exclusively allocated from the part that was assigned to that domain.
The part belonging to the kernel domain is determined by the physical location of the kernel image.\footnote{This is usually at 1MB, although Kernel Address Space Layout Randomization (KASLR) may slightly modify this address according to a limited offset.}
As a result, user and kernel space allocations may be co-located within one bank, but never within adjacent rows.
Different partitioning policies would be possible in theory: for instance, we could confine the kernel to a certain DRAM bank to avoid co-location of user domains within a single bank.
However, this would likely result in a severe increase of memory latency, since reads and writes to a specific memory bank are served by the bank's row buffer.
The benefit of our partitioning policy stems from the fact that we distribute memory belonging to the kernel security domain over multiple banks thereby not negatively impacting performance. 
For our solution towards kernel isolation, we only need to calculate the row index of a page frame.
More specifically, we calculate this index from the physical address (PA) in the following way:
\[Row(PA) := \frac{PA}{PageSize \cdot PagesPerDIMM \cdot DIMMs}\]

Here, we calculate the number of pages per DIMM as $PagesPerDIMM := PagesPerRow \cdot BanksPerRank \cdot RanksPerDIMM$.
Because all possible row indices are present once per bank, this equation determines the row index of the given physical address.\footnote{The default values for DDR3 on x86 are 4K for the page size, 2 pages per row, 8 banks per rank, 2 ranks per DIMM and between 1 one 4 DIMMs per machine. For DDR4 the number of banks per rank was doubled. DDR4 is supported on x86 starting with Intel's Skylake and AMD's Zen architecture.}
We note, that this computation is in line with the available rowhammer exploits \cite{rowhammer-exploit-google} and the reported physical to DRAM mapping recently reverse engineered \cite{drama,rowhammer-vm2}.
Since the row size is the same for all Intel architectures prior to Skylake \cite{rowhammerjs}, our implementation for this policy is applicable to a wide range of system setups, and can be adjusted without introducing major changes to fit other configurations as well.

\noindent \textbf{Isolating User Processes.} To physically isolate individual user processes, we suggest a more flexible partitioning policy, i.e., not requiring user space pages to be confined to a certain part of DRAM.
Instead, we propose to check process pages for conflicting security domains dynamically.
For instance, the kernel could prevent the physical memory allocator from assigning a page to a process, which contains neighboring memory cells in the rows directly above or below, that already belong to another process.
In case such a memory constellation is detected, the buddy allocator would deny handing out the page, otherwise it would be allowed to assign the page to the requesting process.
While this partitioning policy is much more fine-grained than the policy we adopt for isolating the kernel from processes, our implemented mechanism supports such a strategy in principle.
However, implementing this policy is more complex than maintaining a fixed boundary between two domains, since we also need to calculate the channel, rank, and bank index for each page frame in this case.
Hence, an implementation of a process-isolation policy is more specific to the individual hardware setup and would require several different implementations to support additional hardware configurations.

\section{Security Evaluation}
\label{sec:eval:sec}
\begin{table*}[tp]
	\renewcommand{\arraystretch}{1.2}
	\centering
	\begin{tabular}{@{}lllllllllll@{}}
		\toprule
                &                   &                           & \multicolumn{3}{c}{CPU}     & & \multicolumn{4}{c}{RAM}  \\
        \cmidrule{4-6} \cmidrule{8-11} 
        System  & Operating System  & System Model              & Version  & Cores & Speed    & & Size  & Speed    & Manufacturer  & Part number \\
        \midrule
                S1     & Ubuntu 14.04.4 LTS & Dell OptiPlex 7010 0KRC95 & i5-3570  & 4     & 3.40GHz  & & 2x2GB & 1333 MHz & Hynix Hyundai & HMT325U6BFR8C-H9 \\
               &                    &                           &          &       &          & & 1x4GB & 1333 MHz & Corsair       & CMV4GX3M1A1600C11 \\
                S2     & Debian 8.2         & Dell OptiPlex 990         & i7-2600  & 4     & 3.4GHz   & & 2x4GB & 1333 MHz & Samsung       & M378B5273DH0-CH9 \\
                S3     & Kali Linux 2.0     & Lenovo ThinkPad x220      & i5-2520M & 4     & 2.5GHz   & & 2x4GB & 1333 MHz & Samsung       & M471B5273DH0-CH9 \\
		\bottomrule
	\end{tabular}
	\caption{Technical specifications of the vulnerable systems used for our evaluation.}	
	\label{tab:vuln-systems}
\end{table*}

The main goal of our software-based defenses is to protect legacy systems from rowhammer attacks.
We tested the effectiveness of \coolnameBoot and \coolnameGeneric on diverse hardware configurations.
Among these, we identified three hardware configurations, where we observed many reproducible bit flips.
Table~\ref{tab:vuln-systems} lists the exact configurations of the three platforms we use for our evaluation.
Our effectiveness evaluation of \coolnameBoot and \coolnameGeneric is based on two attack scenarios.
For the first scenario we systematically search for reproducible bit flips based on a tool published by Gruss et al.\footnote{\url{https://github.com/IAIK/rowhammerjs/tree/master/native}}
Our second attack scenario leverages a real-world rowhammer exploit\footnote{\url{https://bugs.chromium.org/p/project-zero/issues/detail?id=283}} published by Google's Project Zero.
We compared the outcome of both attacks on our vulnerable systems before and after applying \coolnameBoth.
As shown in Table~\ref{tab:bit-flips}, both tests only succeed when our protections are not in place. Next, we elaborate on the two attack scenarios and their mitigation in more detail.

\begin{table}[tp]
	\renewcommand{\arraystretch}{1.2}
	\centering
	\begin{tabular}{@{}l l l l c l l l@{}}
		\toprule
		\\		&   & \multicolumn{2}{l}{Rowhammer~test} && \multicolumn{3}{l}{Rowhammer Exploit} \\
		&   & \multicolumn{2}{l}{avg. \# reprod. victim pages} && \multicolumn{3}{l}{Success (avg. \# of tries)} \\   
		\cmidrule{3-4} \cmidrule{6-8}
		& Mitigation & None & \coolnameBoot  && None & \coolnameBoot & \coolnameGeneric \\
		\midrule
		S1  &   & 133 & 0 && \cmark (11) & \xmark (4001) & \xmark (3821) \\ 		S2  &   & 31 & 0 && \cmark (42) & \xmark (2897) & \xmark (3096) \\ 		S3  &   & 23 & 0 && \cmark (53) & \xmark (3513) & \xmark (3768) \\ 		\bottomrule
	\end{tabular}
	\caption{Results of our security evaluation. We found that \coolnameBoot and \coolnameGeneric can mitigate rowhammer attacks. We executed the rowhammer test on each system three times and averaged the amount of bit flips.}	
	\label{tab:bit-flips}
\end{table}

\subsection{Rowhammer Testing Tool}\label{sec:rowhammer-test}
We use a slightly modified version of the double-sided rowhammering tool, which is based on the original test by Google's Project Zero~\cite{rowhammer-exploit-google}.
Specifically, we extended the tool to also report the aggressor physical addresses, and adjusted the default size of the fraction of physical memory that is allocated.
The tool scans the system memory for memory cells that are vulnerable to the rowhammer attack. 
To provide comprehensive results, the tool needs to scan the entire memory of the system.
However, investigating the entire memory is hard to achieve in practice since some parts of memory are always allocated by other system components.
These parts are therefore not available to the testing tool, i.e., memory reserved by operating system.
To achieve maximum coverage, the tool allocates a huge fraction of the available memory areas.
However, due to the lazy allocation of Linux the allocated memory is initially not mapped to physical memory.
Hence, each mapped virtual page is accessed at least once, to ensure that the kernel assigns physical pages.
Because user space only has access to the virtual addresses of these mappings, the tool exploits the \texttt{/proc/pagemap} kernel interface to retrieve the physical addresses.
As a result, most of the systems physical memory is allocated to the rowhammering tool. We also discuss on how we achieve almost full coverage in Section~\ref{sec:discussion}.

Afterwards, the tool analyzes the memory in order to identify potential victim and aggressor pages in the physical memory.
As the test uses the double-sided rowhammering approach two aggressor pages must be identified for every potential victim page.
Next, all potential victim pages are challenged for vulnerable bit flips.
For this, the potential victim page is initialized with a fixed bit pattern and ``hammered'' by accessing and flushing the two associated aggressor pages.
This ensures that all of the accesses activate a row in the respective DRAM module.
This process is repeated $10^6$ times.\footnote{This value is the hardcoded default value. Prior research~\cite{rowhammer-study,rowhammer-paper} reported similar numbers.}
Lastly, the potential victim address can be checked for bit flips by comparing its memory content with the fixed pattern bit.
The test outputs a list of addresses for which bit flips have been observed, i.e., a list of victim addresses.

\subsubsection{Preliminary Tests for Vulnerable Systems}
Using the rowhammering testing tool we evaluated our target systems.
In particular, we were interested in systems that yield reproducible bit flips, as only those are relevant for practical rowhammer attacks.
This is because an attacker cannot force the system to allocate page tables at a certain physical position in RAM. In contrast, the attacker sprays the memory with page tables to increase her chance of hitting the desired memory location.

Hence, we configured  the rowhammering tool to only report memory addresses where bit flips can be triggered repeatedly.
We successively confirmed that this list indeed yields reliable bit flips by individually triggering the reported addresses and checking for bit flips within an interval of 10 seconds.
Additionally, we tested the bit flips across reboots through random sampling.

The three systems mentioned in Table~\ref{tab:vuln-systems} are highly susceptible to reproducible bit flips.
Executing the rowhammer test on these three times and rebooting the system after each test run, we found 133 pages with exploitable bit flips for S1, 31 pages for S2, and 23 pages for S3.

\subsubsection{\coolnameBoot}
\label{sec:eval:rowhammertest:bcatt}

We installed \coolnameBoot on our three test systems S1 -- S3.
For each system, we blacklisted the vulnerable physical pages collected in the previous step.
We first verified that the list of vulnerable memory pages was correctly reported to the operating system.
\coolnameBoot adds vulnerable pages as reserved regions to the \texttt{e820 map} (cf.~Section~\ref{sec:impl}). Hence, these pages need to be listed in the \texttt{e820 map} used by the OS.

Next, we executed the rowhammer test on the evaluation systems to verify that it can no longer find vulnerable memory addresses.
We executed the test three times on each system to yield a reasonable coverage.
For all test runs, the rowhammer test failed with an error message when trying to allocate the physical addresses that belongs to a blacklisted page.
As such, all attempts of bit flips on vulnerable pages are prevented by \coolnameBoot.
Thus, \coolnameBoot prevents all previously presented rowhammer attacks~\cite{rowhammer-exploit-google, rowhammer-new-attack, rowhammer-browser, rowhammerjs}.

\subsubsection{\coolnameGeneric}
To install \coolnameGeneric, we patched the Linux kernel of each system to use our modified memory allocator.
Recall that \coolnameGeneric does not aim to prevent bit flips but rather constrain them to a \domain.
Hence, executing the rowhammer test on \coolnameGeneric-hardened systems still locates vulnerable pages. 
However, in the following, we demonstrate based on a real-world exploit that the vulnerable pages are 
not exploitable.

\subsection{Real-world Rowhammer Exploit}
To further demonstrate the effectiveness of our mitigations, we tested \coolnameBoot and \coolnameGeneric against a real-world rowhammer exploit.
The goal of the exploit is to escalate the privileges of the attacker to kernel privileges (i.e., gain root access).
To do so, the exploit leverages rowhammer to manipulate the page tables. Specifically, it aims to manipulate 
the access permission bits for kernel memory, i.e., reconfigure its access permission policy.\footnote{A second option is to manipulate page table entries in such a way that they point to attacker controlled memory thereby allowing the attacker to install new arbitrary memory mappings. The details of this attack option are described in~\cite{rowhammer-exploit-google}.}

To launch the exploit, two conditions need to be satisfied: (1)~a page table entry must be present in a vulnerable row, and (2)~the enclosing aggressor pages must be allocated in attacker-controlled memory.

Since both conditions are not directly controllable by the attacker, the attack proceeds as follows:
the attacker allocates large memory areas. As a result, the operating system needs to create large 
page tables to maintain the newly allocated memory. This in turn increases the probability to satisfy 
the aforementioned conditions, i.e., a page table entry will eventually be allocated to a victim page. 
Due to vast allocation of memory, the attacker also increases her chances that 
aggressor pages are co-located to the victim page.

Once the preconditions are satisfied, the attacker launches the rowhammer attack to induce a bit flip in victim page. Specifically, 
the bit flip modifies the page table entry such that a subtree of the paging hierarchy is under the attacker's control.
Lastly, the attacker modifies the kernel structure that holds the attacker-controlled user process privileges
to elevate her privileges to the superuser root. Since the exploit is probabilistic, it only succeeds in five out of hundred runs. 
Nevertheless, a single successful run allows the attacker to compromise of the entire system.

\subsubsection{\coolnameBoot}
As \coolnameBoot prevents any rowhammer induced bit flips on a system, it also prevents this exploit.
As vulnerable memory pages are not available anymore, the attacker does not succeed in allocating page table entries on vulnerable pages.
Hence, the attacker cannot manipulate the kernel's page tables.

\subsubsection{\coolnameGeneric}
\label{sec:eval:rowhammerexploit:gcatt}
Our generic defense mechanism does not prevent the occurrence of bit flips on a system.
Hence, we have to verify that bit flips cannot affect data of another \domain.
Rowhammer exploits rely on the fact that such a cross domain bit flip is possible, i.e., in the 
case of our exploit it aims to induce a bit flip in the kernel's page table entries.

However, since the exploit by itself is probabilistic, an unsuccessful attempt does not imply the effectiveness of \coolnameGeneric.
As described above, the success rate of the attack is about 5\%.
After deploying \coolnameGeneric on our test systems we repeatedly executed the exploit to minimize the probability of the exploit failing due to the random memory layout rather than due to our protection mechanism.
We automated the process of continuously executing the exploit and ran this test for $48\mathrm{h}$, on all three test systems.
In this time frame. the exploit made on average 3500 attempts of which on average 175 should have succeeded. However, with \coolnameGeneric, none of the attempts 
was successful. Hence, as expected, \coolnameGeneric effectively prevents rowhammer-based exploits.

\label{ref:armeval}
As we have demonstrated, \coolnameGeneric successfully prevents the original attack developed on x86
by physically isolating pages belonging to the kernel from the user-space domain.
In addition to that, the authors of the Drammer exploit~\cite{rowhammer-arm} confirm that \coolnameGeneric prevents their exploit on ARM.
The reason is, that they follow the same strategy as in the original kernel exploit developed by Project Zero, i.e., corrupting page table entries in the kernel from neighboring pages in user space.
Hence, \coolnameGeneric effectively prevents rowhammer exploits on ARM-based mobile platforms as well.

\section{Performance Evaluation}
\label{sec:eval:perf}
One of our main goals is practicability, i.e., inducing negligible performance overhead.
To demonstrate practicability of our two defenses, we thoroughly evaluated the performance and stability impact of \coolnameBoot and \coolnameGeneric on different benchmark and testing suites.
In particular, we used the SPEC CPU2006 benchmark suite~\cite{spec-cpu} to measure the impact on CPU-intensive applications, LMBench3~\cite{lmbench} for measuring the overhead of system operations, and the Phoronix test suite~\cite{phoronix} to measure the overhead for common applications.
We use the Linux Test Project, which aims at stress testing the Linux kernel, to evaluate the stability of our test system after deploying \coolnameBoot and \coolnameGeneric.
We performed all of our performance evaluation on system~S2 (cf. Table~\ref{tab:vuln-systems}).

\subsection{Run-time Overhead}
\begin{table}[t]
    \renewcommand{\arraystretch}{1.2}
    \begin{minipage}[t]{.23\textwidth}
        
        \vspace{0pt}
    	\begin{tabular}[t]{@{}p{43pt} r r@{}}
            \toprule
            \textbf{SPEC2006}   &  \coolnameBoot & \coolnameGeneric \\
            \midrule
            perlbench           & -1.79\% & 0.29\% \\
            bzip2               & -1.81\% & 0.00\% \\
            gcc                 & 0.00\% & -0.71\% \\
            mcf                 & 0.00\% & -1.12\% \\
            gobmk               & -0.46\% & 0.00\% \\
            hmmer               & -1.12\% & 0.23\% \\
            sjeng               & -0.19\% & 0.19\% \\
            libquantum          & 0.88\% & -1.63\% \\
            h264ref             & 0.00\% & 0.00\% \\
            omnetpp             & -0.71\% & -0.28\% \\
            astar               & -1.40\% & -0.45\% \\
            xalan               & 0.00\% & -0.14\% \\
            milc                & 0.00\% & -1.79\% \\
            namd                & 0.40\% & -1.82\% \\
            dealll              & -1.63\% & 0.00\% \\
            soplex              & -0.19\% & 0.00\% \\
            povray              & -0.55\% & -0.46\% \\
            lbm                 & 0.68\% & -1.12\% \\
            sphinx3             & -0.28\% & -0.58\% \\
            \midrule
            \textbf{Mean}       & -0.43\% & -0.49\% \\
            \bottomrule
        \end{tabular}
        
        \vspace{3.25pt}     	
        \begin{tabular}[b]{@{}p{43pt} r r@{}}
            \toprule
            \textbf{Phoronix} &  \coolnameBoot & \coolnameGeneric \\
            \midrule
            IOZone            &  0.57\%  &  0.05\% \\
            Unpack Kernel     & -0.30\%  & -0.50\% \\
            PostMark          & -0.91\%  &  0.92\% \\
            7-Zip             & -0.07\%  &  1.18\% \\
            OpenSSL           & -0.19\%  & -0.22\% \\
            PyBench           & -0.44\%  & -0.59\% \\
            Apache            &  2.49\%  & -0.21\% \\
            PHPBench          &  0.39\%  &  0.35\% \\
            stream            &  0.13\%  &  1.96\% \\
            ramspeed          &  0.00\%  &  0.00\% \\
            cachebench        &  0.02\%  &  0.05\% \\
            \midrule
            \textbf{Mean}     & 0.15\%   &  0.27\% \\
            \bottomrule
        \end{tabular}
    \end{minipage}
    \begin{minipage}[t]{.27\textwidth}
        \centering
        \vspace{0pt}
    	\begin{tabular}{@{}l r r@{}}
            \toprule
            \textbf{LMBench3}       &  \coolnameBoot & \coolnameGeneric \\
            \midrule
            Context Switching:      & & \\
            \quad 2p/0K             & -2.43\% & -2.44\% \\
            \quad 2p/16K            & -2.00\% & 0.00\% \\
            \quad 2p/64K            & 0.00\%  & 2.00\% \\
            \quad 8p/16K            & -1.60\% & -1.73\% \\
            \quad 8p/64K            & -1.90\% & 0.00\% \\
            \quad 16p/16K           & -1.33\% & -1.33\% \\
            \quad 16p/64K           & -0.99\% & 0.99\% \\
            \quad \textbf{Mean}     & -1.58\% & -0,36\% \\
            \midrule
            
            File \& VM Latency:     & & \\
            \quad 0K File Create    & 2.08\% & 0.27\% \\
            \quad 0K File Delete    & 0.98\% & 0.89\% \\
            \quad 10K File Create   & 2.07\% & -0.35\% \\
            \quad 10K File Delete   & 0.51\% & 0.47\% \\
            \quad Mmap Latency      & 0.60\% & -1.81\% \\
            \quad \textbf{Mean}     & 1.13\% & -0,12\% \\
            \midrule
    
            Local Bandwidth:  & & \\
            \quad Pipe              & 0.90\% & 0.18\% \\
            \quad AF UNIX           & 0.31\% & -0.30\% \\
            \quad File Reread       & 1.27\% & -0.38\% \\
            \quad Mmap reread       & 0.00\% & 0.00\% \\
            \quad Bcopy (libc)      & 0.04\% & 0.08\% \\
            \quad Bcopy (hand)      & 0.67\% & 0.34\% \\
            \quad Mem read          & 0.00\% & 0.00\% \\
            \quad Mem write         & 0.94\% & 0.43\% \\
            \quad \textbf{Mean}     & 0.51\% & 0.04\% \\
            \midrule
            
            Memory Latency:   & & \\
            \quad L1 \$             & 0.00\% & 0.00\% \\
            \quad L2 \$             & 0.00\% & 0.00\% \\
            \quad Main mem          & -2.09\% & -2.09\% \\
            \quad Rand mem          & 0.26\% & 1.66\% \\
            \quad \textbf{Mean}     & -0.46\% & 0.11\% \\
            \bottomrule        
        \end{tabular}
    \end{minipage}    \caption{The benchmarking results for SPEC CPU2006, Phoronix, and LMBench3 indicate that \coolnameBoot and \coolnameGeneric induce no measurable performance overhead.
In some cases we observed negative overhead, hence, performance improvements.
However, we attribute such results to measuring inaccuracy.}
    \label{tab:perf}
\end{table}

Table~\ref{tab:perf} summarizes the results of our performance benchmarks.
In general, the SPEC CPU2006 benchmarks measure the impact of system modifications on CPU intensive applications.
Since our mitigations mainly affect the physical memory management, we did not expect a major impact on these benchmarks.
However, since these benchmarks are widely used and well established we included them in our evaluation.\label{rev:measure-inac}
In fact, we observe a minimal performance improvement for \coolnameBoot by $0.43\%$ and \coolnameGeneric by $0.49\%$ which we attribute to measuring inaccuracy.
Such results have been reported before when executing a set of benchmarks for the same system with the exact same configuration and settings. Hence, we conclude that \coolnameBoth does not incur any performance penalty.

LMBench3 is comprised of a number of micro benchmarks which target very specific performance parameters, e.g., memory latency.
For our evaluation, we focused on micro benchmarks that are related to memory performance and excluded networking benchmarks.
Similar to the previous benchmarks, the results fluctuate on average between $-0.4\%$ and $0.11\%$.
Hence, we conclude that both of our mitigations have no measurable impact on specific memory operations.

Finally, we tested the impact of our modifications on the Phoronix benchmarks.
In particular, we selected a subset of benchmarks\footnote{The Phoronix benchmarking suite features a large number of tests which cover different aspects of a system. By selecting a subset of the available tests we do not intend to improve our performance evaluation. On the contrary, we choose a subset of tests that is likely to yield measurable performance overhead, and excluded tests which are unrelated to our modification, e.g., GPU or machine learning benchmarks.} that, on one hand, aim to measure memory performance (IOZone and Stream), and, on the other hand, test the performance of common server applications which usually rely on good memory performance.

To summarize, our rigorous performance evaluation with the help of different benchmarking suites did not yield any measurable overhead.
This makes \coolnameBoot and \coolnameGeneric highly practical mitigations against rowhammer attacks.

\subsection{Memory Overhead}
\label{sec:eval:perf:mem}
\begin{table}[t]
	\renewcommand{\arraystretch}{1.2}
	\centering
	\vspace{0pt}
	\begin{tabular}[t]{@{}l r r r@{}}
		\toprule
		\textbf{System} & Blacklisted pages &  Total pages    & Overhead \\
		\midrule
		S1              & $133$             & $2,097,152$     & $0.0063\%$ \\
		S2              & $31$              & $2,097,152$     & $0.0015\%$ \\
		S3              & $23$              & $2,097,152$     & $0.0011\%$ \\
		
		\bottomrule        
	\end{tabular}
	\caption{Memory overhead for \coolnameBoot.}
	\label{tab:mem-bcatt}
\end{table}

Both, \coolnameBoot and \coolnameGeneric, prevent the operating system from allocating certain physical memory.

For \coolnameBoot the memory overhead depends on the individual system, since all memory pages containing vulnerable memory cells are blacklisted.
As this list is specific per system, we cannot make general statements.
However, we evaluated the memory overhead for our three test systems.
Table~\ref{tab:mem-bcatt} lists the number of blacklisted pages for our test systems S1 -- S3:
the total memory pages available in the system and the fraction of blacklisted pages from the overall memory.
As shown, the memory overhead of \coolnameBoot is at most $0.0063\%$ (<1MB) for our test systems.

The memory overhead of \coolnameGeneric is constant and depends solely on number of memory rows per bank.
Per bank, \coolnameGeneric omits one row to provide isolation between the \domains.
Hence, the memory overhead is $1/\#rows$ ($\#rows$ being rows per bank).
While the number of rows per bank is dependent on the system architecture, it is commonly in the order of $2^{15}$ rows per bank~\footnote{\url{https://lackingrhoticity.blogspot.de/2015/05/how-physical-addresses-map-to-rows-and-banks.html}}, i.e., the overhead is $2^{-15}$ $\hat{=}$ $0,003\%$.

\subsection{Robustness}
\begin{table}[t]
    \renewcommand{\arraystretch}{1.2}
    \centering
    \vspace{0pt}
	\begin{tabular}[t]{@{}l r r r@{}}
        \toprule
        \textbf{Linux Test Project} & Vanilla &  \coolnameBoot  & \coolnameGeneric \\
        \midrule
        clone                       & \cmark  & \cmark          & \cmark \\
        ftruncate                   & \cmark  & \cmark          & \cmark \\
        prctl                       & \cmark  & \cmark          & \cmark \\
        ptrace                      & \cmark  & \cmark          & \cmark \\
        rename                      & \cmark  & \cmark          & \cmark \\
        sched\_prio\_max            & \cmark  & \cmark          & \cmark \\
        sched\_prio\_min            & \cmark  & \cmark          & \cmark \\
        mmstress                    & \cmark  & \cmark          & \cmark \\
        shmt                        & \xmark  & \xmark          & \xmark \\
        vhangup                     & \xmark  & \xmark          & \xmark \\
        ioctl                       & \xmark  & \xmark          & \xmark \\
       \bottomrule        
    \end{tabular}
    \caption{Result for individual stress tests from the Linux Test Project.}
    \label{tab:robustness}
\end{table}

Our mitigations restrict the operating systems access to the physical memory.
To ensure that this has no effect on the overall stability, we performed numerous stress tests with the help of the Linux Test Project (LTP)~\cite{ltp}.
These tests are designed to cause problems for the operating system.
We first run these tests on a vanilla Debian 8.2 installation to receive a baseline for the evaluation of \coolnameBoot and \coolnameGeneric.
We summarize our results in Table~\ref{tab:robustness}, and report no deviations for our mitigations compared to the baseline.
Further, we also did not encounter any problems during the execution of the other benchmarks.
Thus, we conclude that \coolnameBoot and \coolnameGeneric do not affect the stability of the protected system.

\section{Discussion}
\label{sec:disc}
Both of our prototype implementations target Linux-based systems.
Linux is open-source allowing us to implement our defenses. Further, all publicly available rowhammer attacks target this operating system.
\coolnameBoot already protects all operating systems that can be loaded with the commodity bootloader GRUB2. 
Further, \coolnameGeneric can be easily ported to memory allocators deployed in other operating systems.
In this section, we discuss in detail the generality and coverage aspects of our software-based defenses against rowhammer.

\subsection{Generality Aspects of \coolnameBoot} \label{sec:gen-bcatt}
\label{sec:disc:bcatt}
Extending the GRUB2 bootloader to blacklist vulnerable memory allows us to protect all Linux-based systems, OpenBSD, and Xen~\cite{xen}.
Xen is an open-source hypervisor that is widely deployed on production systems.
Similar to the Linux kernel, Xen receives the list of available and reserved memory through GRUB2.
Hence, the current implementation of \coolnameBoot can defeat recently presented rowhammer attacks in a cloud scenario~\cite{rowhammer-vm1, rowhammer-vm2}.

Unfortunately, GRUB2 is not capable of directly booting Windows systems.
Instead, in order to boot Windows, GRUB2 starts the native bootloader of the Windows installation.
This process is called \emph{chainloading}.
GRUB2 does not pass the list of unavailable memory addresses to the Windows bootloader, i.e., the Windows components discover the memory layout by themselves.
To make \coolnameBoot available to Windows three approaches are possible:
(1)~Microsoft extends its bootloader to include the modifications we applied to GRUB2.
(2)~The BIOS vendor adapts \coolnameBoot and provides a modified \texttt{e820} which includes vulnerable memory.
(3)~GRUB2 hooks into the \texttt{e820} BIOS interrupt allowing \coolnameBoot to add vulnerable memory when the Windows bootloader searches for usable memory.

In the current implementation of \coolnameBoot we do not support UEFI.
However, since UEFI provides a similar mechanism to the \texttt{e820 map}, there are no obstacles to add support for UEFI in \coolnameBoot.

\subsection{Generality Aspects of \coolnameGeneric} \label{sec:gen-gcatt}
\label{sec:disc:gcatt}
With our current implementation of \coolnameGeneric we focus on isolating the user and kernel memory to prevent privilege escalation attacks, because this is the most practical and single publicly available attack vector.
However, as described in Section~\ref{sec:impl:processisolation}, we are currently looking into isolating memory of user processes, i.e., assigning a \domain per process.
Although we are not aware of any cross-process rowhammer attacks we anticipate to see such attacks in the future as an alternative to kernel-based privilege escalation attacks. Further, these attacks are conceptually similar to cross-VM attacks which have been already demonstrated recently~\cite{rowhammer-vm1}.

By isolating the physical memory of different processes, \coolnameGeneric can stop such attacks (cf. Section~\ref{sec:impl:processisolation}).
Similarly, when implementing \coolnameGeneric in the hypervisor, physical memory of different virtual machines (VMs) can be isolated.

\subsubsection{Fine-Grained In-Process Isolation}
\begin{figure}[tp!]
	\centering
	\includegraphics[width=\linewidth]{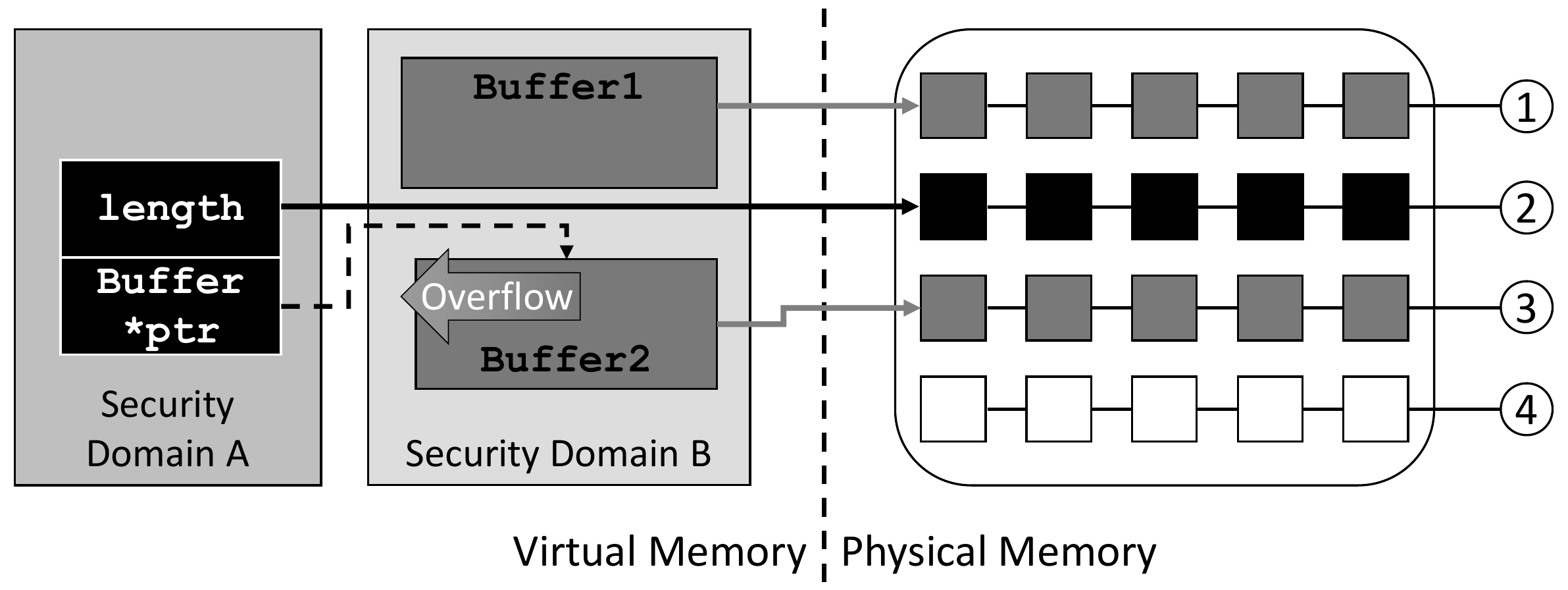}
		\caption{Virtual memory isolation vs. physical memory isolation. }
	\label{fig:ieattack}
\end{figure}

In fact, our design also allows an fine-grained isolation \emph{within} a process.
Separation within a single process has many security applications, e.g., application sandboxes, or isolating attacker-controllable data.
As an example, in 2014, Microsoft improved the heap allocator for Internet Explorer which separates the heap and objects on the heap into two parts.\footnote{https://technet.microsoft.com/en-us/library/security/ms14-035.aspx}
As shown in Figure~\ref{fig:ieattack}, data objects in virtual memory  are divided into meta data and payload data.
The payload data (\texttt{Buffer1} and \texttt{Buffer2}) are attacker-controlled data.
The meta data (\texttt{length} and \texttt{*ptr}) are usually isolated from the attacker.
In the example from Figure~\ref{fig:ieattack}, the array object is comprised of a memory buffer (\texttt{Buffer2}), a pointer to the buffer (\texttt{Buffer *ptr}) and a field for the length (\texttt{length}). 
On a monolithic heap the meta data (\texttt{length} and \texttt{*ptr}) would be adjacent with the corresponding buffer (\texttt{Buffer2}). 
In this case, the attacker can overwrite the meta data by overflowing \texttt{Buffer2}.

The separated heap, as shown in Figure~\ref{fig:ieattack}, prevents this attack, as an overflow in \texttt{Buffer2} would only affect other attacker-controlled data (e.g., \texttt{Buffer1}).
However, the separation of meta data and payload data exists only in the virtual memory space.
In the physical memory space meta data and payload data can still be stored in adjacent locations:
Figure~\ref{fig:ieattack} shows a possible mapping of the data into physical memory.

Given this setting, the attacker can modify the meta data by leveraging rowhammer.
By ``hammering'' \texttt{Buffer1}  and \texttt{Buffer2}, she can launch a double-sided rowhammer attack.
As a result, she can modify data belonging to \domain A by accessing data of \domain B~\cite{rowhammerjs,rowhammer-browser}.

To prevent such attacks, the \domain isolation needs to be expanded into the physical memory space. 
\coolnameGeneric allows the creation of arbitrary \domains.
However, \coolnameGeneric is implemented in the kernel while the semantics of the security domains (the separated heaps) are only known to the application.
Hence, to enable applications to have isolation in the physical memory space this information must be provided to the physical memory allocator in the kernel.
This can be achieved by extending the memory allocation interface of the kernel (e.g., \texttt{malloc}) with an additional parameter allowing an application to indicate which security domain should be assigned to the newly allocated memory.

\subsubsection{Applying \coolnameGeneric to Mobile Systems}
The rowhammer attack is not limited to x86-based systems, but has been recently shown to also affect the ARM platform~\cite{rowhammer-arm}.
The ARM architecture is predominant in mobile systems, and used in many smartphones and tablets.
As \coolnameGeneric is not dependent on any x86 specific properties, it can be easily adapted for ARM based systems.
We demonstrate this by applying our extended physical memory allocator to the Android kernel for Nexus devices in version 4.4.
Since there are no major deviations in the implementation of the physical page allocator of the kernel between
Android and stock Linux kernel, we did not encounter any obstacles during the port.
\label{ref:armimpl}

\subsection{Coverage of Rowhammer Testing Tool} \label{sec:coverage}
\label{sec:discussion}

\coolnameBoot relies on the fact that all vulnerable memory pages can be identified and blacklisted. 
In general, the problem of discovering vulnerable bits is orthogonal to the enforcement mechanism by \coolnameBoot.
For our prototype implementation and evaluation, we use the rowhammer testing tool developed and published by Google~\cite{rowhammer-memtest}. However, there are two factors that impact the comprehensiveness of identifying vulnerable bits: spatial coverage and temporal coverage.
Since identifying vulnerable memory is not required for \coolnameGeneric, we focus on \coolnameBoot for the following discussion on this aspect.

\subsubsection{Spatial Coverage Aspects}
\begin{figure*}[tp!]
	\centering
	\includegraphics[width=0.75\linewidth]{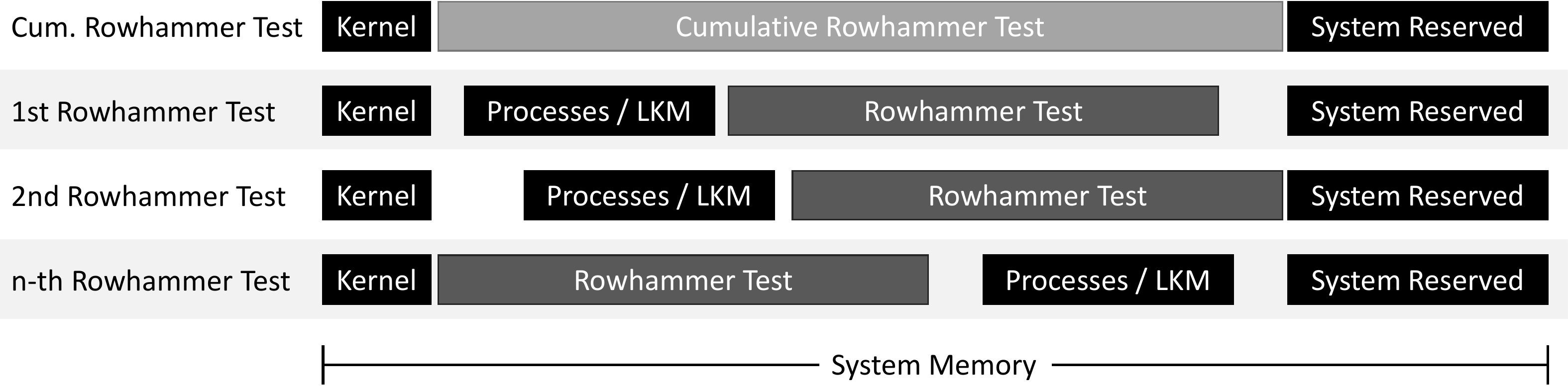}
		\caption{Memory allocations of rowhammer testing tools across reboots. Each individual run can test only a fraction of the system memory. By combining repetitive tests the coverage can be maximized.}
	\label{fig:coverage}
	\vspace{-2em}
\end{figure*}

The memory coverage of the rowhammer testing tool is limited to the physical memory it can access.
Since the rowhammer testing tool executes as an ordinary process on the system, it has no influence on the physical memory assigned when allocating memory.
In particular, there are parts of physical memory which will never be assigned to the rowhammer test tool, as some memory is reserved by other systems entities, e.g., memory reserved by the kernel.
Figure~\ref{fig:coverage} shows an abstract view of the memory assignment during multiple runs of the rowhammer testing tool. Note that the kernel memory as well as the reserved memory can never be assigned to a process.
Hence, the rowhammer testing tool can never test these regions for vulnerable bits.

Additionally, the rowhammer testing tool has to share the system's memory with other processes.
Figure~\ref{fig:coverage} shows that the rowhammer testing tool cannot allocate all memory since other processes (and other dynamic components like loadable kernel modules -- LKM) also require memory.
However, since these memory allocations are dynamic they can easily change among reboots and runs.
Figure~\ref{fig:coverage} shows that the coverage of the rowhammer testing tool for each individual run is limited.
However, the \emph{cumulative} coverage of the rowhammer testing tool achieves coverage of the entire process' allocatable memory.
Lastly, even if the rowhammer testing tool cannot detect all vulnerable memory bits, it still can detect all bits that are exploitable from the attacker's point of view, i.e., the attacker leverages the same techniques as the rowhammer testing tool to detect vulnerable bits.

\subsubsection{Temporal Coverage Aspects}
Rowhammer is a novel attack technique. To the best of our knowledge, there are no studies on the system's susceptibility to rowhammer over a large time frame.
However, one could easily imagine that memory cells could wear-off over time and cells which have been unaffected before become vulnerable.
Since there is no data available whether this can happen or at which rate cells can become vulnerable, no definitive statements can be made at the time of writing. 
In general, if cells become vulnerable over time, the detection of vulnerable pages would need to be repeated periodically for \coolnameBoot.
As mentioned before, these updates can be fully automated to run without requiring user interaction.
\label{ref:temporal}
Moreover, an attacker cannot simply utilize newly appearing victim pages, but would have to repeat the memory profiling for the targeted machine as well.

\subsection{Single-sided Rowhammer Attacks}
From our detailed description in Section~\ref{sec:highlvl:gcatt} one can easily follow that both of our proposed solutions can defeat all known rowhammer attacks in general, and single-sided rowhammer attacks~\cite{rowhammer-arm} in particular. 
In contrast to double-sided rowhammer attacks (see Figure~\ref{fig:dram2}), single-sided rowhammer attacks relax the adversary's capabilities by requiring that the attacker has control over only one row adjacent to the victim memory row.
However, as shown in Figure~\ref{fig_design} \coolnameBoot will prevent the system from allocating memory that is vulnerable to rowhammer attacks.
Hence, even if the system contains memory that is vulnerable to single-sided rowhammer, \coolnameBoot ensures that this memory remains unused throughout the entire run time of the system.
As described in more detail in Section~\ref{sec:highlvl:gcatt}, \coolnameGeneric isolates different security domains in the physical memory.
In particular, it ensures that different security domains are separated by at least one buffer row that is never used by the system.
This means that the single-sided rowhammer adversary can only flip bits in own memory (that it already controls), or flip bits in buffer rows.

\subsection{Benchmarks Selection}
We selected our benchmarks to be comparable to the related literature.
Moreover, we have done evaluations that go beyond those in the existing work to provide additional insight.
Hereby, we considered different evaluation aspects:
We executed SPEC CPU2006 to verify that our changes to the operating system impose no overhead of user-mode applications.
Further, SPEC CPU2006 is the most common benchmark in the field of memory-corruption defenses, hence, our solutions can be compared to the related work.
LMBench3 is specifically designed to evaluate the performance of common system operations, and used by the Linux kernel developers to test whether changes to the kernel affect the performance.
As such LMBench3 includes many tests.
For our evaluation we included those benchmarks that perform memory operations and are relevant for our defenses.
Indeed we have also tested all other memory-related benchmarks in LMBench3 and observed no measurable overhead (0.2\%) as well.
Finally, we selected a number of common applications from the Phoronix test suite as macro benchmarks, as well as the \emph{pts/memory} tests which are designed to measure the RAM and cache performance.
For all our benchmarks we did not observe any measurable overhead (see Table~\ref{tab:perf}).

\section{Related Work}
\label{sec:relatedwork}
In this section, we provide an overview of existing rowhammer attack techniques, their evolution, and proposed defenses. Thereafter, we discuss the shortcomings of existing work on mitigating rowhammer attacks and compare them to our two software-based defenses.  

\subsection{Rowhammer Attacks}
Kim et al.~\cite{rowhammer-paper} were the first to conduct experiments and analyze the effect of bit flipping due to repeated memory reads.
They found that this vulnerability can be exploited on Intel and AMD-based systems. Their results show that over 85\% of the analyzed DRAM modules are vulnerable.
The authors highlight the impact on memory isolation, but they do not provide any practical attack.
Seaborn and Dullien~\cite{rowhammer-exploit-google} published the first practical rowhammer-based privilege-escalation attacks using the x86 \texttt{clflush} instruction.
In their first attack, they use rowhammer to escape the Native Client (NaCl)~\cite{nacl} sandbox.
NaCl aims to safely execute native applications by 3rd-party developers in the browser.
Using rowhammer malicious developers can escape the sandbox, and achieve remote code execution on the target system.
With their second attack Seaborn and Dullien utilize rowhammer to compromise the kernel from an unprivileged user-mode application.
Combined with the first attack, the attacker can remotely compromise the kernel without exploiting any software vulnerabilities. 
To compromise the kernel, the attacker first fills the physical memory with page-table entries by allocating a large amount of memory.
Next, the attacker uses rowhammer to flip a bit in memory.
Since the physical memory is filled with page-table entries, there is a high probability that an individual page-table entry is modified by the bit flip in a way that enables the attacker to access other page-table entries, modify arbitrary (kernel) memory, and eventually completely compromise the system.
Qiao and Seaborn~\cite{rowhammer-new-attack} implemented a rowhammer attack with the x86 \texttt{movnti} instruction.
Since the \texttt{memcpy} function of \texttt{libc}~--~which is linked to nearly all C programs~--~utilizes the \texttt{movnti} instruction, the attacker can exploit the rowhammer bug with code-reuse attack techniques~\cite{rop-shacham}.
Hence, the attacker is not required to inject her own code but can reuse existing code to conduct the attack.
Aweke et al.~\cite{anvil} showed how to execute the rowhammer attack without using any special instruction (e.g., \texttt{clflush} and \texttt{movnti}).
The authors use a specific memory-access pattern that forces the CPU to evict certain cache sets in a fast and reliable way.
They also concluded that a higher refresh rate for the memory would not stop rowhammer attacks.
Gruss et al.~\cite{rowhammerjs} demonstrated that rowhammer can be launched from JavaScript.
Specifically, they were able to launch an attack against the page tables in a recent Firefox version.
Similar to Seaborn and Dullien's exploit this attack is mitigated by \coolnameGeneric.
Later, Bosman et al.~\cite{rowhammer-browser} extended this work by exploiting the memory deduplication feature of Windows~10 to create counterfeit JavaScript objects, and corrupting these objects through rowhammer to gain arbitrary read/write access within the browser.
In their follow-up work, Razavi et al.~\cite{rowhammer-vm1} applied the same attack technique to compromise cryptographic (private) keys of co-located virtual machines.
Concurrently, Xiao et al.~\cite{rowhammer-vm2} presented another cross virtual machine attack where they use rowhammer to manipulate page-table entries of Xen.
Further, they presented a methodology to automatically reverse engineer the relationship between physical addresses and rows and banks.
Independently, Pessl et al.~\cite{drama} also presented a methodology to reverse engineer this relationship.
Based on their findings, they demonstrated cross-CPU rowhammer attacks, and practical attacks on DDR4.
Van der Veen et al.~\cite{rowhammer-arm} recently demonstrated how to adapt the rowhammer exploit to escalate privileges in Android on smartphones.
Since the authors use the same exploitation strategy of Seaborn and Dullien, \coolnameGeneric can successfully prevent this privilege escalation attack.
While the authors conclude that it is challenging to mitigate rowhammer in software, we present two viable implementations that can mitigate practical rowhammer attacks.

Note that all these attacks require memory belonging to a higher-privileged domain (e.g., kernel) to be physically co-located to memory that is under the attacker's control.
Since our defenses prevent direct co-location, we mitigate these rowhammer attacks.

\subsection{Defenses Against Rowhammer}
Kim et al.~\cite{rowhammer-paper} present a number of possible mitigation strategies.
Most of their solutions involve changes to the hardware, i.e., improved chips, refreshing rows more frequently, or error-correcting code memory.
However, these solutions are not very practical: the production of improved chips requires an improved design, and a new manufacturing process which would be costly, and hence, is unlikely to be implemented.
The idea behind refreshing the rows more frequently (every 32ms instead of 64ms) is that the attacker needs to hammer rows many times to destabilize an adjacent memory cell which eventually causes the bit flip. 
Hence, refreshing (stabilizing) rows more frequently could prevent attacks because the attacker would not have enough time to destabilize individual memory cells.
\label{ref:refreshrate}
Nevertheless, Aweke et al.~\cite{anvil} were able to conduct a rowhammer attack within 32ms.
Therefore, a higher refresh rate alone cannot be considered as an effective countermeasure against rowhammer.
Error-correcting code (ECC) memory is able to detect and correct single-bit errors.
As observed by Kim et al.~\cite{rowhammer-paper} rowhammer can induce multiple bit flips which cannot be detected by ECC memory.
Further, ECC memory has an additional space overhead of around 12\% and is more expensive than usual DRAM, therefore it is rarely used.

Kim et al.~\cite{rowhammer-paper} also suggest to disable faulty rows similar to our \coolnameBoot that we present in Section~\ref{sec:highlvl:bcatt}. However, they provide no implementation and evaluation. In contrast, we present the full implementation and evaluation of \coolnameBoot.  
While the authors concluded that this strategy is not practical, we demonstrate that it is an efficient and practical solution that effectively prevents rowhammer attacks as a short-term solution. 
Kim et al. suggest to use probabilistic adjacent row activation (PARA) to mitigate rowhammer attacks.
As the name suggests, reading from a row will trigger an activation of adjacent rows with a low probability.
During the attack, the malicious rows are activated many times. Hence, with high probability the victim row gets refreshed (stabilized) during the attack.
The main advantage of this approach is its low performance overhead. However, it requires changes to the memory controller.
Thus, PARA is not suited to protect legacy systems.

To the best of our knowledge Aweke et al.~\cite{anvil} proposed the only other software-based mitigation against rowhammer.
Their mitigation, coined ANVIL, uses performance counters to detect high cache-eviction rates which serves as an indicator of rowhammer attacks~\cite{anvil}.
However, this defense strategy has three disadvantages:
(1)~it requires the CPU to feature performance counters. In contrast, our defenses do not rely on any special hardware features.
(2)~ANVIL's worst case run-time overhead for SPEC CPU2006 is 8\%, whereas our worst case overhead is 0.29\% (see Table~\ref{tab:perf}).
(3)~ANVIL is a heuristic-based approach. Hence, it naturally suffers from false positives (although the FP rate is below 1\% on average).
In contrast, we provide deterministic approaches that are guaranteed to stop rowhammer attacks.

\section{Conclusion}
Rowhammer is a hardware fault, triggered by software, allowing the attacker to flip bits in physical memory and undermine CPU-enforced memory access control. Recently, researchers have demonstrated the power and consequences of rowhammer attacks by breaking the isolation between virtual machines, user and kernel mode, and even enabling traditional memory-corruption attacks in the browser.

In this paper we introduce two practical and effective defenses against rowhammer, \coolnameBoot and \coolnameGeneric. 
\coolnameBoot extends the bootloader to blacklist vulnerable memory which completely prevents the attacker from inducing bit flips through rowhammer.
\coolnameGeneric, on the other hand, is a generic mitigation that tolerates rowhammer attacks by dividing the physical memory into \domains, and limiting rowhammer-induced bit flips to the attacker-controlled \domain.

Our detailed evaluation of \coolnameBoot and \coolnameGeneric verifies that both schemes prevent all known rowhammer attacks.
Further, our mitigation schemes do not affect the run-time performance or the stability of the system.

\section*{Acknowledgment}
The authors thank Simon Schmitt for sacrificing his personal laptop to the cause of science, and Victor van der Veen, Daniel Gruss and Kevin Borgolte for their feedback.

This work was supported in part by the German Science Foundation (project S2,  CRC 1119 CROSSING), the European Union's Seventh Framework Programme (609611, PRACTICE), and the German Federal Ministry of Education and Research within CRISP.

\bibliographystyle{abbrv}

\end{document}